%% file: main.tex
\documentclass[sigconf]{acmart} 

\usepackage{graphicx}
\usepackage{balance}  

\usepackage{graphics}
\usepackage{xspace}
\usepackage[ruled, vlined, linesnumbered]{algorithm2e} 

\usepackage{booktabs}
\usepackage{soul}
\usepackage{balance}

\usepackage{float}
\usepackage{algorithmicx, algpseudocode}
\usepackage{amsmath}
\usepackage{amsfonts}
\usepackage{amsbsy}
\usepackage{amsthm}
\usepackage[]{caption}
\usepackage{multirow}
\usepackage[mathscr]{eucal} 
\usepackage{mathtools}
\usepackage{verbatim}
\usepackage{tabularx}
\usepackage{enumitem}
\usepackage{array}
\usepackage{soul}
\usepackage{dsfont}
\usepackage{url}
\usepackage{bbm}
\usepackage{pifont}
\usepackage{xcolor,colortbl}
\usepackage[inkscapeformat=png]{svg}
\usepackage{makecell}

\usepackage{contour}
\usepackage{ulem}
\contourlength{0.8pt}

\newcommand{\myuline}[1]{%
  \uline{\phantom{#1}}%
  \llap{\contour{white}{#1}}%
}

\usepackage{xspace}

\newcommand{\measure}{\textsc{HyperTrans}\xspace}
\newcommand{\naivemeasure}{\textsc{Naive-HyperTrans}\xspace}
\newcommand{\fastmeasure}{\textsc{Fast-HyperTrans}\xspace}

\newcommand{\generator}{\textsc{THera}\xspace}
\newcommand{\naivegenerator}{\textsc{Naive-THera}\xspace}
\newcommand{\random}{\textsc{Null}\xspace}

\usepackage[center]{subfigure}

\newtheorem{definition}{\textbf{Definition}}
\newtheorem{theorem}{\textbf{Theorem}}

\newtheorem{proposition}{\textbf{Proposition}}

\newcommand\modify[1]{\textcolor{blue}{#1}}
\newtheorem{axiombox}{\textbf{Axiom}}
\newtheorem{observation}{\textbf{Observation}}

\newcommand{\newaxiomtwo}[1]{%
  \newtheorem{subaxiomtwo#1}{\textbf{Axiom}}%
  \expandafter\renewcommand\csname thesubaxiomtwo#1\endcsname{#1\Alph{subaxiomtwo#1}}%
}

\newaxiomtwo{1}
\newaxiomtwo{2}

\newcommand{\newaxiomthree}[1]{%
  \newtheorem{subaxiomthree#1}{\textbf{Axiom}}%
  \expandafter\renewcommand\csname thesubaxiomthree#1\endcsname{#1\Alph{subaxiomthree#1}}%
}

\newaxiomthree{1}
\newaxiomthree{2}
\newaxiomthree{3}

\newcommand{\newaxiomtwoa}[1]{%
  \newtheorem{subaxiomtwoa#1}{\textbf{Axiom}}%
  \expandafter\renewcommand\csname thesubaxiomtwoa#1\endcsname{#1\Alph{subaxiomtwoa#1}}%
}

\newaxiomtwoa{1}
\newaxiomtwoa{2}

\newcommand{\newaxiomthreea}[1]{%
  \newtheorem{subaxiomthreea#1}{\textbf{Axiom}}%
  \expandafter\renewcommand\csname thesubaxiomthreea#1\endcsname{#1\Alph{subaxiomthreea#1}}%
}

\newaxiomthreea{1}
\newaxiomthreea{2}
\newaxiomthreea{3}

\definecolor{newgreen}{rgb}{0.0, 0.5, 0.0}
\definecolor{newred}{rgb}{0.81,0.1,0.26}
\definecolor{babyblue}{rgb}{0.54, 0.81, 0.94}
\definecolor{negcol}{rgb}{0.96, 0.76, 0.76}
\definecolor{poscol}{rgb}{0.63, 0.79, 0.95}
\definecolor{sunwoogreen}{rgb}{0.66, 0.89, 0.63}
\definecolor{sunwoogreentwo}{rgb}{0.53, 0.81, 0.98}
\definecolor{dodgerblue}{rgb}{0.12, 0.56, 1.0}
\definecolor{crimson}{rgb}{0.86, 0.08, 0.24}
\definecolor{limegreen}{rgb}{0.2, 0.8, 0.2}
\definecolor{backredcolor}{rgb}{0.91, 0.45, 0.32}

\newcommand{\posc}{\cellcolor{poscol}}  
\newcommand{\negc}{\cellcolor{negcol}}  
\newcommand{\goat}{\cellcolor{sunwoogreen}}  
\newcommand{\goattwo}{\cellcolor{sunwoogreentwo}}  

\newcommand{\yes}{\textcolor{newgreen}{\textbf{\ding{52}}}}
\newcommand{\no}{\textcolor{newred}{\textbf{\ding{55}}}}
\DeclarePairedDelimiter{\ceil}{\lceil}{\rceil}
\let\oldnl\nl
\newcommand{\nlnonumber}{\renewcommand{\nl}{\let\nl\oldnl}}

\AtBeginDocument{%
  \providecommand\BibTeX{{%
    \normalfont B\kern-0.5em{\scshape i\kern-0.25em b}\kern-0.8em\TeX}}}

\copyrightyear{2023}
\acmYear{2023}
\setcopyright{acmlicensed}\acmConference[KDD '23]{Proceedings of the 29th ACM SIGKDD Conference on Knowledge Discovery and Data Mining}{August 6--10, 2023}{Long Beach, CA, USA}
\acmBooktitle{Proceedings of the 29th ACM SIGKDD Conference on Knowledge Discovery and Data Mining (KDD '23), August 6--10, 2023, Long Beach, CA, USA}
\acmPrice{15.00}
\acmDOI{10.1145/3580305.3599382}
\acmISBN{979-8-4007-0103-0/23/08}

\ccsdesc[500]{Information systems~Data mining}
\ccsdesc[500]{Information systems~Social networks}
\keywords{Hypergraph; Group Interaction; Transitivity; Generator}

\begin{document}

    \title{How Transitive Are Real-World Group Interactions? - Measurement and Reproduction}
    
	
	\author{Sunwoo Kim}
	\affiliation{%
    	\institution{KAIST}
            \city{}
            \country{}
	}
	\email{kswoo97@kaist.ac.kr}

        \author{Fanchen Bu}
	\affiliation{%
		\institution{KAIST}
            \city{}
            \country{}
	}
	\email{boqvezen97@kaist.ac.kr}

        \author{Minyoung Choe}
	\affiliation{%
		\institution{KAIST}
            \city{}
            \country{}
	}
	\email{minyoung.choe@kaist.ac.kr}
    
	\author{Jaemin Yoo}
	\affiliation{%
            \institution{Carnegie Mellon University}
            \city{}
            \country{}
	}
	\email{jaeminyoo@cmu.edu}
	
	\author{Kijung Shin}
	\affiliation{%
		\institution{KAIST}
            \city{}
            \country{}
	}
	\email{kijungs@kaist.ac.kr}

    \begin{abstract}
		\input{000abstract.tex}
	\end{abstract}
	
	\input{dfn.tex}
	\maketitle

        \section{Introduction}
	\label{sec:intro}
	\input{001Introduction.tex}

	\section{Concepts and Axioms}
	\label{sec:conceptandaxiom}
	\input{002PrelimAxiom.tex}

	\section{Proposed Measure and Algorithm}
	\label{sec:measure}
	\input{003AxiomMeasure.tex}
    
        \section{Datasets and Patterns}
	\label{sec:observation}
	\input{004Observations.tex}
        \section{Pattern-Preserving Generator}
	\label{sec:generator}
	\input{005Generator.tex}
        \vspace{-2mm}

        \section{Related Work}
	\label{sec:relatedwork}
	\input{006RelatedWork.tex}

        \section{Conclusion}
	\label{sec:conclusion}
	\input{007Conclusion.tex}

        \vspace{1mm}
        {\small \smallsection{Acknowledgements} This work was supported by National Research Foundation of Korea (NRF) grant funded by the Korea government (MSIT) (No. NRF-2020R1C1C1008296) and Institute of Information \& Communications Technology Planning \& Evaluation (IITP) grant funded by the Korea government (MSIT) (No. 2022-0-00871, Development of AI Autonomy and Knowledge Enhancement for AI Agent Collaboration) (No. 2019-0-00075, Artificial Intelligence Graduate School Program (KAIST)).
        }

        \bibliographystyle{ACM-Reference-Format}
        \balance
	\bibliography{BIB/ref}

        
        \clearpage
        \appendix 
        
        \section{Appendix: Explanation of axioms}
        \input{100AppendixMotivation}
        \label{sec:motivation}
        
        \section{Appendix: Analyses of \measure variants (B7-9)}

\input{101BaselineUsecase}
        \label{sec:variantbaseline}
        
        
        \section{Appendix: Experimental Details}
        \input{102AppendixExperiment.tex}
        \label{sec:expsetting}

	

	
\end{document}

%% file: 000Abstract.tex
Many real-world interactions (e.g., researcher collaborations and email communication) occur among multiple entities.
These group interactions are naturally modeled as hypergraphs.
In graphs, transitivity is helpful to understand the connections between node pairs sharing a neighbor, and it has extensive applications in various domains.
Hypergraphs, an extension of graphs, are designed to represent group relations. 
However, to the best of our knowledge, there has been no examination regarding the transitivity of real-world group interactions.
In this work, we investigate the transitivity of group interactions in real-world hypergraphs.
We first suggest intuitive axioms as necessary characteristics of hypergraph transitivity measures. 
Then, we propose a principled hypergraph transitivity measure \measure, which satisfies all the proposed axioms, with a fast computation algorithm \fastmeasure.
After that, we analyze the transitivity patterns in real-world hypergraphs distinguished from those in random hypergraphs.
Lastly, we propose a scalable hypergraph generator \generator. 
It reproduces the observed transitivity patterns by leveraging community structures, which are pervasive in real-world hypergraphs.
Our code and datasets are available at \url{https://github.com/kswoo97/hypertrans}.


%% file: dfn.tex
\newcommand\red[1]{\textcolor{red}{#1}}
\newcommand\blue[1]{\textcolor{blue}{#1}}
\newcommand\gray[1]{\textcolor{gray}{#1}}

\definecolor{peace}{RGB}{228, 26, 28}
\definecolor{love}{RGB}{55, 126, 184}
\definecolor{joy}{RGB}{77, 175, 74}
\definecolor{kindness}{RGB}{152, 78, 163}

\newcommand\peace[1]{\textcolor{peace}{#1}}
\newcommand\love[1]{\textcolor{love}{#1}}
\newcommand\joy[1]{\textcolor{joy}{#1}}
\newcommand\kindness[1]{\textcolor{kindness}{#1}}

\newcommand\kijung[1]{\textcolor{peace}{[Kijung: #1]}}
\newcommand\fanchen[1]{\textcolor{love}{[Fanchen: #1]}}
\newcommand\minyoung[1]{\textcolor{joy}{[Minyoung: #1]}}
\newcommand\sunwoo[1]{\textcolor{kindness}{[Sunwoo: #1]}}

\newcommand{\smallsection}[1]{{\vspace{0.02in} \noindent {{\myuline{\smash{\bf #1:}}}}}}
\newtheorem{obs}{\textbf{Observation}}
\newtheorem{defn}{\textbf{Definition}}
\newtheorem{thm}{\textbf{Theorem}}
\newtheorem{axm}{\textbf{Axiom}}
\newtheorem{lma}{\textbf{Lemma}}
\newtheorem{cor}{\textbf{Corollary}}
\newtheorem{problem}{\textbf{Problem}}
\newtheorem{pro}{\textbf{Problem}}
\newtheorem{remark}{\textbf{Remark}}

\newcommand\und[1]{\underline{#1}}

\newcommand{\SM}{\mathcal{M}}%
\newcommand{\SG}{\mathcal{G}}%
\newcommand{\SSS}{\mathcal{S}}%
\newcommand{\SV}{\mathcal{V}}%
\newcommand{\SE}{\mathcal{E}}%
\newcommand{\SL}{\mathcal{L}}%
\newcommand{\SGH}{\mathcal{\hat{G}}}%
\newcommand{\SEH}{\mathcal{\hat{E}}}%
\newcommand{\SVH}{\mathcal{\hat{V}}}%
\newcommand{\SEVH}{\SE({\SVH})}%
\newcommand{\SVEH}{\SV({\SEH})}%
\newcommand{\SD}{\mathcal{D}}%

\newcommand{\fhat}{\hat{f}}%
\newcommand{\flargehat}{\hat{F}}%
\newcommand{\yhat}{\hat{y}}%

\newcommand{\Ghat}{\SGH=(\SVH,\SEH)}%
\newcommand{\Glong}{\SG=(\SV,\SE)}%

\newcommand{\cmark}{\ding{51}}%
\newcommand{\xmark}{\ding{55}}%

\newcommand{\method}{\textsc{WHATsNet}\xspace}
\newcommand{\attention}{\textsc{ATT}\xspace}
\newcommand{\aggregation}{\textsc{AGG}\xspace}
\newcommand{\MultiheadAtt}{\textsc{MAB}\xspace}
\newcommand{\WithinAttentionFull}{within attention\xspace}
\newcommand{\WithinAttention}{\textsc{WithinATT}\xspace}
\newcommand{\PE}{\textsc{WithinOrderPE}\xspace}

\newcommand{\methodgrid}{\textsc{MiDaS-Grid}\xspace}
\newcommand{\methodauto}{\textsc{MiDaS}\xspace}

\definecolor{myred}{RGB}{195, 79, 82}
\definecolor{mygreen}{RGB}{86, 167 104}
\definecolor{myblue}{RGB}{74, 113 175}

\newcommand{\bigcell}[2]{\begin{tabular}{@{}#1@{}}#2\end{tabular}}

\let\oldnl\nl
\newcommand{\nonl}{\renewcommand{\nl}{\let\nl\oldnl}}

%% file: 001Introduction.tex
\begin{figure}[t]
    \centering
    \subfigure[Example Substances Dataset]{\includegraphics[width=0.1935\textwidth]{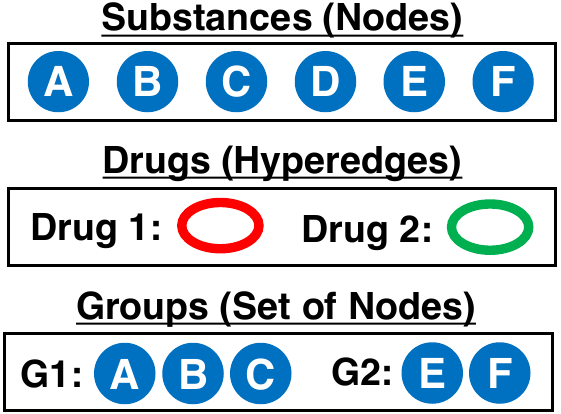}}
    \hspace{3mm}
    \subfigure[Hypergraph]{\includegraphics[width=0.153\textwidth]{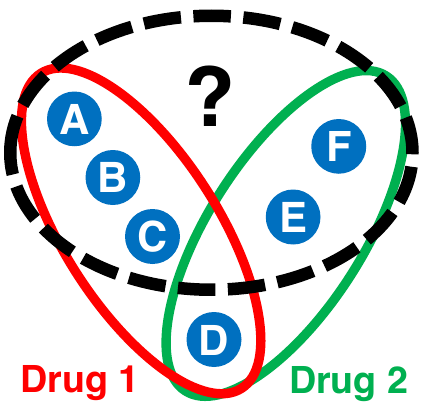}}
    \vspace{-1mm}
    \caption{\label{fig:overview} {Two drugs and their six constituent components (substances) modeled as a hypergraph.}}
\end{figure}

Going beyond pairwise relations, real-world interactions often involve multiple entities.
For example, scholars collaborate on research, substances interact to form drugs, and people engage in communication on social media.
\textbf{\textit{Hypergraphs}}, which are a generalization of graphs, are a commonly used data structure for modeling group interactions. 
Each hypergraph consists of a node set and a hyperedge set, where each hyperedge is a set that can include any number of nodes. 
In Figure~\ref{fig:overview}, 
we provide an example of modeling drugs and their constituent components as a hypergraph.

\begin{figure*}[t!]
  \centering
  \vspace{-3.5mm}
  \begin{tabular}{c | c | c | c}
  \begin{subfigure}[\textsc{\textbf{Axiom 1}}]
      {\includegraphics[width=0.22\textwidth]{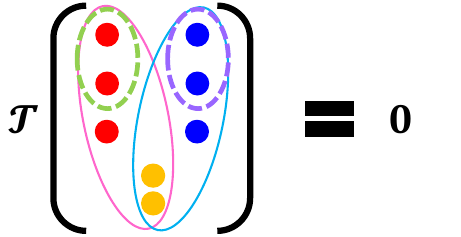}}      
    \end{subfigure}&
  \begin{subfigure}[\textsc{\textbf{Axiom 2: Case 1}}]
      {\includegraphics[width=0.22\textwidth]{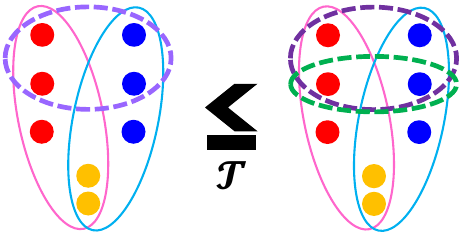}}      
    \end{subfigure}&
    \begin{subfigure}[\textsc{\textbf{Axiom 2: Case 2}}]
      {\includegraphics[width=0.22\textwidth]{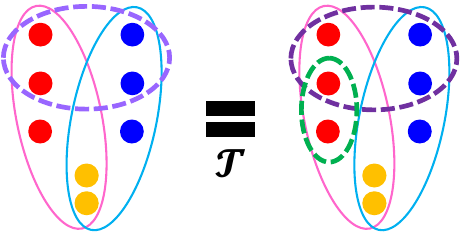}}      
    \end{subfigure}&
    \begin{subfigure}[\textsc{\textbf{Axiom 2: Case 3}}]
      {\includegraphics[width=0.22\textwidth]{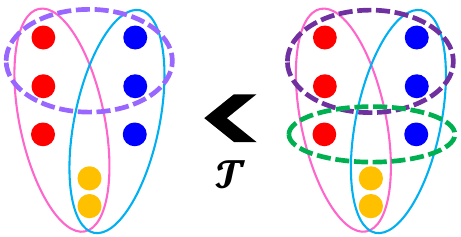}}      
    \end{subfigure} \\
    \midrule
   \begin{subfigure}[\textsc{\textbf{Axiom 3: Case 1}}]
      {\includegraphics[width=0.22\textwidth]{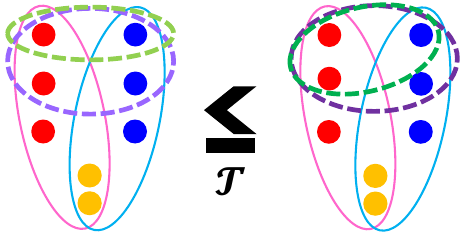}}      
    \end{subfigure}&
    \begin{subfigure}[\textsc{\textbf{Axiom 3: Case 2}}]
      {\includegraphics[width=0.22\textwidth]{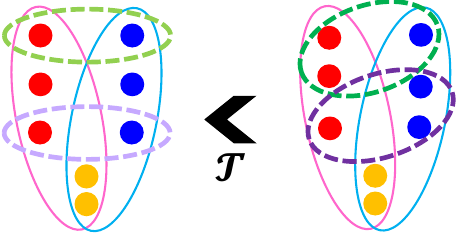}}      
    \end{subfigure}&
    \begin{subfigure}[\textsc{\textbf{Axiom 4}}]
      {\includegraphics[width=0.22\textwidth]{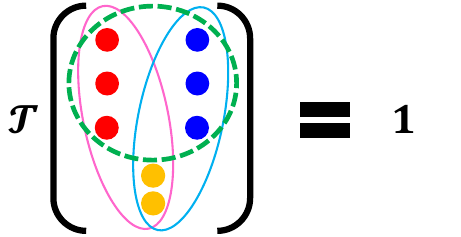}}      
    \end{subfigure}&
    \begin{subfigure}
      {\includegraphics[width=0.22\textwidth]{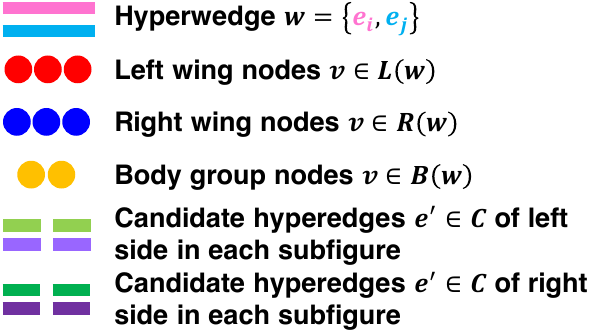}}      
    \end{subfigure}
  \end{tabular}   
  \caption{Examples for \textsc{Axiom}~\ref{ax:min_hw_trans}-\ref{ax:max_hw_trans}.}
  \label{fig:overallaxiom}
\end{figure*}

\textit{\textbf{Transitivity}}, also known as a clustering coefficient, is a {measure of the likelihood} of two neighbors of a node in a graph being adjacent~\cite{wasserman1994social, watts1998collective, newman2001random}. 
As a key graph statistic, transitivity has been used in diverse fields, including neuroscience~\cite{masuda2018clustering, hsu2018applying, loeffler2020topological},
{bioinformatics \cite{gallagher2013clustering},} and finance~\cite{tabak2014directed, cerqueti2021systemic}, for  various applications, including web analysis~\cite{kutzkov2013streaming, becchetti2010efficient}, and link prediction~\cite{chen2019application, wu2016link}. 

{Measuring the \textbf{transitivity of group interactions} is of potential importance in applications where (a) group interactions are prevalent and (b) transitivity provides essential information.
Prominent examples of such applications include protein interaction analysis \cite{gallagher2013clustering} and financial risk management \cite{cerqueti2021systemic}.}


Although various properties of real-world group interactions have been examined, their transitivity is still underexplored.
Several attempts to measure transitivity in hypergraphs~\cite{gallagher2013clustering, zhou2011properties, klamt2009hypergraphs, torres2021and, estrada2005complex, behague2021iterated} 
{essentially focused on pairwise relations between individual nodes, spec.,} whether two neighbors of a node are included in the same hyperedge(s) together, {overlooking higher-order information beyond pairs.}
Furthermore, some measures can only quantify the overall transitivity in a hypergraph but cannot quantify local transitivity, e.g., transitivity around each hyperedge or each node.

{Measuring the transitivity of group interactions presents new challenges. For instance, when considering two hyperedges (e.g., Drugs 1 and 2 in Figure~\ref{fig:overview}), quantifying the transitivity around them in a systematic manner requires addressing (a) the possibility that multiple hyperedges may overlap with both hyperedges and (b) the possibility that they intersect with different parts of the two hyperedges (e.g., $\{A, B, E\}$ and $\{A, C, F\}$ in Figure~\ref{fig:overview}). These possibilities arise due to the inherent characteristics of group interactions.} 


In this work, we investigate the transitivity patterns of 12 real-world hypergraphs and develop a generator to reproduce the observed realistic transitivity patterns.
Our contributions toward these goals are summarized as follows:
\begin{enumerate}[wide, labelindent=5pt]
    \item \textbf{Axioms and a principled transitivity measure:} We propose seven intuitive properties that a proper hypergraph transitivity measure should satisfy, and we formalize them into axioms. 
    Then, we propose \measure, a transitivity measure that satisfies all the axioms (while all existing measures fail to do so), with a fast computation algorithm \fastmeasure.
    \item \textbf{Observations on real-world hypergraphs:} We analyze real-world hypergraphs' transitivity patterns, and we show that these patterns are different from those of random hypergraphs generated by a null model.
    The analyses are conducted at the levels of hypergraphs, hyperwedges, nodes, and hyperedges.
    \item \textbf{Generator:} We propose \generator, a scalable hypergraph generator  that reproduces the transitivity patterns on real-world hypergraphs. 
    \generator utilizes a mechanism based on community structures, offering insights into the underlying foundation of transitivity of real-world group interactions
\end{enumerate}

In Section~\ref{sec:conceptandaxiom}, we provide some preliminaries and the axioms. 
In Section~\ref{sec:measure}, we introduce \measure, a principled hypergraph transitivity measure. 
In Section~\ref{sec:observation}, we explore the transitivity patterns of real-world hypergraphs. 
In Section~\ref{sec:generator}, we present \generator, a hypergraph generator that successfully reproduces the observed patterns.
In Section~\ref{sec:conclusion}, we give the conclusion of our work.

%% file: 002PrelimAxiom.tex
In this section, we introduce several basic concepts related to hypergraph transitivity. 
Then, we propose seven axioms on the necessary characteristics of a desirable hypergraph transitivity measure.
See Table~\ref{tab:notations} for the frequently-used symbols.

\subsection{Basic Concepts}\label{subsec:basic_concepts}
\smallsection{Preliminaries} 
A \textit{\textbf{hypergraph}} $G = (V,E)$ consists of a node set $V = \{v_{1} , \cdots , v_{|V|}\}$ and a hyperedge set $E = \{e_{1} , \cdots , e_{|E|}\}$, where each \textit{\textbf{hyperedge}} is a set of nodes i.e., $e_{i} \subseteq V, \forall i \in \{1,\cdots |E|\}$.
{A \textbf{\textit{hyperwedge}} is defined as a pair of intersecting hyperedges each of which is not a subset of the other.
That is, if $w = \{e_i, e_j\}$ is hyperwedge, then
 $e_i \cap e_j \neq \emptyset$, $e_i \not\subseteq e_j$, and $e_i \not\subseteq e_j$ hold.}
Thus, the set of hyperwedges $W = W(G)$ in $G$ is defined as
\[
W(G) = \{\{e_{i}, e_{j}\} \in \binom{E}{2}: e_{i} \cap e_{j} \neq \emptyset \wedge e_{i} \not \subseteq e_{j} \wedge e_{j} \not \subseteq e_{i} \}.
\]
For each $w = \{e_{i},e_{j}\} \in W$, we define 
the \textit{left wing} of $w$ as $L(w) = e_i \setminus e_j$, and we define
the \textit{right wing} of $w$ as $R(w) = e_j \setminus e_i$. {We give the two wings different names for ease of presentation, but the two wings are essentially symmetric since $w = \{e_{i},e_{j}\} = \{e_{j},e_{i}\}$.}
We also define the \textit{body group} of $w$ as $B(w) = e_{i} \cap e_{j}$ and define $P(w)$, the set of all the possible \textit{pair interactions} between the nodes in $L(w)$ and those in $R(w)$, i.e.,
\begin{equation}\label{eq:wedgespairinter}
P(w) = \{\{v'_{1}, v'_{2}\} : v'_{1} \in L(w), v'_{2} \in R(w)\}.
\end{equation}

\smallsection{Hypergraph transitivity measures}
We will define measures of hypergraph transitivity on two different levels: a \textit{hyperwedge-level measure} $\mathcal{T}$ and a \textit{hypergraph-level measure} $T$.

At the hyperwedge-level, we aim to assess the group interactions between the nodes in the two \textit{disjoint} wings, $L(w)$ and $R(w)$.
For each hyperwedge $w$, let $\Omega(w)$ denote the set of \textit{overlapping hyperedges} w.r.t $w$, where an overlapping hyperedge is a hyperedge that intersects both wings of $w$.
Formally, $\Omega(w) = \{e \in E: e \cap L(w) \neq \emptyset \land e \cap R(w) \neq \emptyset\}$.
Note that for each $w$, there may exist multiple overlapping hyperedges,
i.e., it is possible that $|\Omega(w)| > 1$;
while in pairwise graphs there may exist at most one overlapping edge (i.e., when $\lvert e\vert  = 2, \forall e \in E$ \text{ then } $\lvert \Omega(w) \vert \in \{0, 1\}, \forall w \in W$). 
Therefore, computing transitivity in hypergraphs is a \textit{nontrivial} extension of the counterpart in graphs, where \textbf{the multiplicity of overlapping hyperedges} should be taken into account.
Finally, given a hypergraph $G = (V, E)$, a \textit{target hyperwedge} $w$, and a non-empty \textit{candidate set} $C \subseteq E$,
we use $\mathcal{T}(w, C; G)$ to denote the hyperwedge-level transitivity measure of $w$ in $G$ w.r.t $C$,
where a candidate set consists of \textit{candidate hyperedges} that contribute to the transitivity of $w$.
When $C = E$, we may simply use $\mathcal{T}(w; G)$ to denote $\mathcal{T}(w, E; G)$.
Moreover, when the context is clear, we may omit the input hypergraph $G$ and use $\mathcal{T}(w, C)$ to denote $\mathcal{T}(w, C; G)$ (and thus we use $\mathcal{T}(w)$ to denote $\mathcal{T}(w, E; G)$).
For hypergraph-level measure $T$, we follow a common way to scale from local transitivity to global transitivity~\cite{watts1998collective}, which is an average of all hyperwedge transitivity in a given hypergraph (i.e., $T(G) = \sum_{w \in W(G)} \mathcal{T}(w;G)/\lvert W(G)\vert$).

\begin{table}[t!]
    \vspace{-3mm}
	\caption{\label{tab:notations}Frequently-used symbols.}
        \resizebox{\linewidth}{!}{
		\begin{tabular}{c|l}
			\toprule
			\textbf{Notation} & \textbf{Definition}\\
			\midrule
			$G = (V,E)$ & a hypergraph with nodes $V$ and hyperedges $E$\\
                $W = W(G)$ & the set of hyperwedges in a hypergraph $G$ \\
                $T(G)$ & the transitivity of a hypergraph $G$\\
			$L(w), R(w)$ & the left wing and the right wing of a hyperwedge $w$ \\
                $P(w)$ & the set of possible pair interactions between the two wings of $w$ (Eq~\eqref{eq:wedgespairinter}) \\
                $B(w)$ & the body group of a hyperwedge $w$ \\
                $\mathcal{T}(w, C; G)$ & the transitivity of a target hyperwedge $w$ in $G$ w.r.t a candidate set $C$\\
			$f$ & a group interaction function \\
			\bottomrule
		\end{tabular}}
\end{table}

\subsection{Axioms and baseline measures}\label{subsec:intuitiveaxiom}
\smallsection{Axioms}
What properties must a well-defined and intuitive hypergraph transitivity measure possess?
We propose seven axioms to formally describe such desirable properties, including five hyperwedge-level axioms (Axioms~\ref{ax:min_hw_trans}-\ref{ax:hw_bounded}), where we assume that the input hypergraph $G$ is fixed,
and two hypergraph-level axioms 
(Axioms~\ref{ax:reduce_pairwise_graph} and \ref{ax:hg_bounded}).
The motivation and necessity of each axiom are given in Appendix~\ref{subsec:axiommoti}.
In Figure~\ref{fig:overallaxiom}, we provide examples for Axioms~\ref{ax:min_hw_trans}-\ref{ax:max_hw_trans}.

In five hyperwedge-level axioms (Axiom~\ref{ax:min_hw_trans}-\ref{ax:hw_bounded}), we assert that they should hold for each hyperwedge $w \in W(G)$. 
We use $C$ and $C'$ to denote two different candidate sets, and their conditions will be explicitly mentioned in each axiom.
\begin{axiombox}[Minimum hyperwedge transitivity]\label{ax:min_hw_trans}
A hyperwedge transitivity of $w$ is globally minimized if and only if there is no candidate hyperedge in $C$ being an overlapping hyperedge (see Figure~\ref{fig:overallaxiom}(a)).
Formally, $\mathcal{T}(w, C) = 0$ (see Axiom~\ref{ax:hw_bounded}) $\Leftrightarrow$ $C \cap \Omega(w) = \emptyset$.
\end{axiombox}

\begin{axiombox}\label{ax:incre_changes_cands}
In this axiom, we discuss how hyperwedge transitivity should change in different situations when we \textbf{include more hyperedges} in the candidate set $C$.
\end{axiombox}
\begin{enumerate}[wide, label=\textsc{\textbf{Case \arabic*}}:, labelindent=10pt]
    \item \textit{(General) 
    Whenever more hyperedges are included in the candidate set $C$, $w$'s transitivity remains the same or increases (see Figure~\ref{fig:overallaxiom}(b)).
    Formally, $C \subseteq C' \subseteq E \Rightarrow \mathcal{T}(w , C) \leq \mathcal{T}(w , C')$.}
    
    \item \textit{(Only non-overlapping) 
    When only non-overlapping hyperedges 
    are further included in $C$, $w$'s transitivity remains the same (see Figure~\ref{fig:overallaxiom}(c)).
    Formally, $(C \subseteq C' \subseteq E) \wedge \Big((C' \setminus C) \cap \Omega(w) = \emptyset\Big)\Rightarrow \mathcal{T}(w , C) = \mathcal{T}(w , C')$.}
    
    \item {\textit{(More interactions covered in total)
    When some hyperedges are further included in $C$ so that more interactions in $P(w)$ are covered, $w$'s transitivity strictly increases (see Figure~\ref{fig:overallaxiom}(d)).
    Formally, $(C \subseteq C' \subseteq E) \wedge \left(\exists e' \in C': (\binom{e'}{2} \setminus \bigcup_{e \in C} \binom{e}{2}) \cap P(w) \neq \emptyset\right)
     \Rightarrow \mathcal{T}(w , C) <  \mathcal{T}(w , C')$.}}
\end{enumerate}

\begin{axiombox}\label{ax:incre_changes_replacement}
In this axiom, we discuss how hyperwedge transitivity should change in different situations when some candidate hyperedges in $C$ are \textbf{enlarged with wing-nodes}, i.e., replaced by their supersets where the new nodes are from the two wings.\footnote{Formally, for each hyperwedge $w$, a candidate hyperedge $e$ is \textit{enlarged with wing-nodes} (to $e'$) if and only if $e \subseteq e'$ with $\emptyset \neq (e' \setminus e) \subseteq (L(w) \cup R(w))$.}
\end{axiombox}
\begin{enumerate}[wide, label=\textsc{\textbf{Case \arabic*}}:, labelindent=10pt]
    \item \textit{(General) 
    When each $e \in C$ is either kept the same or enlarged with wing-nodes, $w$'s transitivity remains the same or increases (see Figure~\ref{fig:overallaxiom}(e)).
    Formally, $\Big(\exists$ bijection $g: C \to C'$ s.t 
    $\big(e \subseteq g(e) \subseteq (e \cup L(w) \cup R(w)), \forall e \in C\big)\Big)
    \Rightarrow \mathcal{T}(w, C) \leq \mathcal{T}(w, C')$.}
    
    \item \textit{(Each candidate more interaction-covering)
    When each $e \in C$ is enlarged with wing-nodes so that it covers more interactions in $P(w)$,
    $w$'s transitivity strictly increases (see Figure~\ref{fig:overallaxiom}(f)).
    Formally, $\Big(\exists$ bijection $g: C \to C'$ s.t
    $\big(e \subsetneq g(e) \subseteq (e \cup L(w) \cup R(w))
    \wedge (\binom{g(e)}{2} \setminus \binom{e}{2} ) \cap P(w) \neq \emptyset, \forall e \in C\big)\Big) 
    \Rightarrow \mathcal{T}(w, C) < \mathcal{T}(w, C')$.}
\end{enumerate}

\begin{remark}
    Axiom~\ref{ax:incre_changes_replacement} assumes a \textbf{bijection}, {ensuring that} the enlarged hyperedges do not become equivalent {to any} other hyperedges.
\end{remark}

\begin{axiombox}[Maximum hyperwedge transitivity]\label{ax:max_hw_trans}
When hyperwedge transitivity of $w$ is globally maximized, there exists at least one $e \in C$ including all the nodes of two wings $L(w)$ and $R(w)$ (see Figure~\ref{fig:overallaxiom}(g)). 
Formally, $\mathcal{T}(w, C) = 1 \Rightarrow \exists e \in C$ s.t $L(w) \cup R(w) \subseteq e$.\footnote{See Appendix~\ref{subsec:whynotheotherwayround} for the discussion on the converse statement.}
\end{axiombox}

\begin{remark}
    Since axioms focus on \textbf{group interaction}, \textsc{Axiom}~\ref{ax:max_hw_trans} implies that all elements in $P(w)$ should co-exist in a single hyperedge.
\end{remark}

\begin{axiombox}[Boundedness of hyperwedge transitivity]\label{ax:hw_bounded}
    A hyperwedge transitivity function $\mathcal{T}$ should be bounded.
    WLOG, we assume that the value is bounded within $[0, 1]$, i.e.,
    $\mathcal{T}(w, C) \in [0, 1], \forall w \in W, \forall C \in 2^{E} \setminus \{\emptyset\}$.    
\end{axiombox}

We now propose two hypergraph-level axioms.
\begin{axiombox}[Reducibility to graph transitivity]\label{ax:reduce_pairwise_graph}
When the input hypergraph $G = (V, E)$ is a pairwise graph, i.e., $|e| = 2, \forall e \in E$, the hypergraph transitivity $T(G)$ should be equal to (i.e., is reduced to) the graph transitivity~\cite{newman2001random} of $G$.
\end{axiombox}

\begin{axiombox}[Boundedness of hypergraph transitivity]\label{ax:hg_bounded}
A hypergraph transitivity function $T$ should be bounded.
WLOG,
$T(G) \in [0, 1]$, for every \text{hypergraph $G$}.
\end{axiombox}

\begingroup
\begin{table}[t!]
    \vspace{-3mm}
	\caption{\label{tab:axiomsatisfy}Only \measure {satisfies} all the axioms.}
        \renewcommand{\arraystretch}{0.9}
	\scalebox{0.77}{
		\begin{tabular}{l |c  c  c  c  c  c c}
			\toprule
			\multirow{2}{*}{\textbf{Measure}} & \multicolumn{7}{c}{\textbf{Axioms}}\\
                & \textbf{1} & \textbf{2} & \textbf{3} & \textbf{4} & \textbf{5} & \textbf{6} & \textbf{7}\\
			\midrule
                \textbf{B1} (Jaccard index)                            & \no & \no & \no & \no & \yes & \yes & \yes\\
                \textbf{B2} (Ratio of covered interacations)                & \yes & \yes & \no & \no & \yes & \yes & \yes\\
                \textbf{B3} (Klamt et al.~\cite{klamt2009hypergraphs})         & \yes & \no & \no & \no & \yes & \yes & \yes\\
                \textbf{B4} (Torres et al.~\cite{torres2021and})               & \yes & \yes & \no & \no & \yes & \yes & \yes\\
                \textbf{B5} (Gallager et al.~\cite{gallagher2013clustering} A) & \no & \no & \no & \no & \yes & \yes & \yes\\
                \textbf{B6} (Gallager et al.~\cite{gallagher2013clustering} B) & \no & \no & \no & \no & \yes & \no  & \yes\\
                \textbf{B7} (\measure-mean)                  & \yes & \no & \yes & \yes & \yes & \yes & \yes\\
                \textbf{B8} (\measure-non-$P(w)$)                  & \yes & \no & \yes & \yes & \yes & \yes & \yes\\
                \textbf{B9} (\measure-unnormalized)                  & \yes & \yes & \yes & \yes & \no & \yes & \no\\
                \midrule 
                Proposed: \measure                                             & \yes & \yes & \yes & \yes & \yes & \yes & \yes\\
			\bottomrule
		\end{tabular}}
\end{table}
\endgroup

\subsection{Baseline Measures}\label{subsec:baselines}

We present several baseline measures, {all of which} violate at least one of the axioms presented in Section~\ref{subsec:intuitiveaxiom}. 
The baseline measures are intuitive quantities, extended from existing measures, or the variants of our finally proposed measure.
Here, we briefly describe each baseline method, and details (e.g., formulae) are provided in the online appendix~\cite{github}.

First, \textbf{B1} and \textbf{B2} are two simple and intuitive measures.

\smallsection{B1. Jaccard similarity} \textbf{B1} computes the Jaccard similarity between (1) the union of all candidate hyperedges in $C$ and (2) the union of the two wings of $w$. 

\smallsection{B2. Ratio of covered interactions} \textbf{B2} computes the ratio of pair interactions in $P(w)$ (Eq~\eqref{eq:wedgespairinter})
that are covered by (included in) the candidate hyperedges.

Baseline methods \textbf{B3-6} are extensions of existing hypergraph transitivity measures~\cite{torres2021and, gallagher2013clustering, klamt2009hypergraphs}.
Since no existing measures were defined at the hyperwedge level, we adapt the concept of local measures (e.g., the local clustering coefficient of a node) to extend the existing measures to the hyperwedge level.

\smallsection{B3. Klamt et al.~\cite{klamt2009hypergraphs}} \textbf{B3} computes the proportion of candidate hyperedges that intersect with both wings out of those that intersect with at least one wing.  

\smallsection{B4. Torres et al.~\cite{torres2021and}} \textbf{B4} computes the proportion of {wing-nodes (i.e., nodes that belong to a wing of the target hyperwedge)} that {are} in the same candidate hyperedge with a node in the other wing.

\smallsection{B5. Gallagher et al. A~\cite{gallagher2013clustering}} 
\textbf{B5} computes the proportion {of pairs of wing-nodes} that co-exist in a candidate hyperedge that is disjoint with the body group $B(w)$ out of {all pairs of wing-nodes.}

\smallsection{B6. Gallagher et al. B~\cite{gallagher2013clustering}} \textbf{B6} is similar to \textbf{B5}, except that the candidate hyperedges that intersect with the body group $B(w)$ are considered, instead of those disjoint with $B(w)$.

In Section~\ref{subsec:theory}, we  provide three more baseline measures 
\textbf{B7} (Eq~\eqref{eq:baselinewomax}), \textbf{B8} (Eq~\eqref{eq:baselineonlymax}), and \textbf{B9} (Eq~\eqref{eq:baselinenonorm}) which are variants of the proposed measure~\measure (Eq~(\ref{eq:hypertrans})).
As shown in Theorem~\ref{thm:axiomviolation} (see also Table~\ref{tab:axiomsatisfy}), 
all the baseline measures violate at least one {of the proposed axioms},
while \measure satisfies them all.

\begin{theorem}[Unconformity of baseline measures]\label{thm:axiomviolation} 
Each baseline measure (\textbf{B1-9}) violates at least one {among Axioms~\ref{ax:min_hw_trans}-\ref{ax:hg_bounded}.}
\end{theorem}

\begin{proof}
See the online appendix~\cite{github}.
\end{proof}

%% file: 003AxiomMeasure.tex
In this section, we introduce a principled hypergraph transitivity measure \textbf{\measure} 
(\myuline{\textbf{Hyper}}graph \myuline{\textbf{Trans}}itivity), which satisfies all the proposed axioms.
In addition, we present a fast and exact computation algorithm \textbf{\fastmeasure}.

\subsection{Proposed Measure: \measure}\label{subsec:transpa}

\smallsection{Definition and intuitions}
We first provide the formal definition of the proposed hyperwedge-level transitivity, \measure:
\begin{equation}\label{eq:hypertrans}
\mathcal{T}(w, C; f) =
\sum\nolimits_{\{v'_{1}, v'_{2}\} \in P(w)} 
\frac{\max_{e \in C} f(w, e) \mathds{1}[v'_{1}, v'_{2} \in e]} 
{\lvert P(w) \vert },
\end{equation}
where 
$\mathds{1}$ is an indicator function,
$P(w)$ has been defined in Eq~\eqref{eq:wedgespairinter},
and {$f : W \times E \mapsto \mathbb{R}$} is a group interaction function that will be specified later.
Informally, given a hyperwedge $w$ and a candidate hyperedge $e \in C$, $f(w, e)$ computes the contribution of $e$ to the interaction between $L(w)$ and $R(w)$. 
For each pair $\{v'_1, v'_2\} \in P(w)$, \measure chooses the candidate hyperedges that include $\{v'_1, v'_2\}$.
By using chosen candidate hyperedges, \measure assigns the \textit{"interaction score"} to the $\{v'_1, v'_2\}$.
Since there may be multiple candidate hyperedges covering (i.e., containing) $v'_1$ and $v'_2$, \measure uses the candidate hyperedge with the highest group interaction function value $f(w, e)$ among the selected candidate hyperedges and regards it as the interaction score of the $\{v'_1 , v'_2\}$ (i.e., $\max_{e \in C} f(w, e) \times \mathds{1}[v'_{1}, v'_{2} \in e]$).
At last, \measure computes the average value of the interaction scores over all the pairwise interactions $\{v'_1, v'_2\} \in P(w)$.
Note that $\mathds{1}[v'_{1}, v'_{2}]$ ensures that only the hyperedges covering $v'_1$ and $v'_2$ are considered.
{More rationales for the designs} in \measure ($\max$, $P(w)$, and divided by $\lvert P(w) \vert$) will be elaborated (see, e.g., \textbf{B7-9} in Section~\ref{subsec:theory} and Theorem~\ref{thm:axiomviolation}).

Following {some related works} on graphs~\cite{watts1998collective, barrat2004architecture},
we define the global transitivity, \textbf{hypergraph transitivity measure} $T$, as the mean of local transitivity values.
Formally, given a hypergraph $G$,
\begin{equation}\label{eq:globalhypertrans}
    T(G) = \frac{1}{\lvert W\vert } \sum\nolimits_{w \in W} \mathcal{T}(w),
\end{equation}
{where} $\mathcal{T}(w)=\mathcal{T}(w,E)$ (see Section~\ref{subsec:basic_concepts}).

\smallsection{Group interaction function $f$}
As mentioned above, the group interaction function $f$ computes the contribution of a candidate hyperedge to the interaction between the two wings of a hyperwedge.
In general, any function $f$ that reasonably represents such contributions can be used.
An intuitive definition of $f$ can be 
\begin{equation}~\label{eq:nopenalize}
f(w, e) = \frac{\lvert L(w) \cap e \vert \times \lvert R(w) \cap e \vert}{\lvert L(w)\vert \times \lvert R(w)\vert },
\end{equation}
which computes the proportion of interactions in $P(w)$ that are covered by the candidate hyperedge $e$.
One may also want to additionally penalize the inclusion of external nodes (i.e., $v \not\in (L(w) \cup R(w)$) in the candidate hyperedge.
In such cases, $f$ can be defined as
\begin{equation}~\label{eq:penalizeall}
f(w, e) = \frac{\lvert L(w) \cap e\vert \times \lvert R(w) \cap e \vert}{\lvert L(w) \cup (e \setminus R(w))\vert \times \lvert R(w) \cup (e \setminus L(w)) \vert }.
\end{equation}

\subsection{Theoretical Analysis}\label{subsec:theory}
Despite the flexibility, the final transitivity measure $\mathcal{T}$ (Eq~\eqref{eq:hypertrans}) should satisfy the necessary characteristics in Axioms~\ref{ax:min_hw_trans}-\ref{ax:hg_bounded}.
Below, we discuss the conditions that ensure $\mathcal{T}$ satisfies these axioms.
\begin{definition}
    A group interaction function $f$ is \textit{\textbf{good}}, if $f$ satisfies the following six properties for each $w$ and $e$:
    \begin{enumerate}[wide, labelindent=5pt]
        \item $f : (w, e) \in [0, 1], \forall w \in W(G), e \in E$.\label{prop:bound}
        \item $e \in \Omega(w) \Rightarrow f(w,e) > 0$.\label{prop:notzero}
        \item $f(w,e) = 1 \Rightarrow L(w) \cup R(w) \subseteq e$.\label{prop:perfect2edge}
        \item $L(w) \cup R(w) = e \Rightarrow f(w,e) = 1$.\label{prop:edge2perfect}
        \item $e \subseteq e' \subseteq (e \cup L(w) \cup R(w)) \Rightarrow f(w,e) \leq f(w, e')$.\label{prop:grow1}
        \item $e \subsetneq e' \subseteq (e \cup L(w) \cup R(w)) \wedge e' \in \Omega(w) \Rightarrow f(w,e) < f(w, e')$.\label{prop:grow2}
    \end{enumerate}
\end{definition}


    
    
    
    
    

\begin{theorem}[Soundness of \measure]\label{thm:measureaxiom}
\measure (Eq~(\ref{eq:hypertrans})) with a good group interaction score function $f$ 
satisfies \textsc{Axiom 1-7}.
\end{theorem}
\begin{proof}{
See the online appendix~\cite{github}.
}
\end{proof}
By Theorem~\ref{thm:measureaxiom}, one can use any \textit{good} $f$ to ensure that $\mathcal{T}$ satisfies all the axioms.
Throughout {the remaining parts of the paper}, we use the definition of $f$ in Eq~\eqref{eq:penalizeall}, which is good as shown below.
\begin{proposition}~\label{prop:choice}
The function $f$ defined in Eq~\eqref{eq:penalizeall} is good.
\end{proposition}
\begin{proof}{
See the online appendix~\cite{github}.
}
\end{proof}

Below, we provide three variants of \measure, \textbf{B7-9}, as three additional baseline measures.
By Theorem~\ref{thm:axiomviolation} (see also Table~\ref{tab:axiomsatisfy}), \textbf{B7-9} fail to satisfy the axioms,
validating the necessity of the designs in \measure.
Specifically, \textbf{B7} shows the necessity of the $\max$ function, 
\textbf{B8} emphasizes the importance of considering all interactions in $P(w)$, 
and \textbf{B9} demonstrates the significance of dividing by $\lvert P(w) \vert$. 
Limitations of \textbf{B7-9} are described {in detail} in Appendix~\ref{sec:variantbaseline}.

\smallsection{B7. \measure-mean} \textbf{B7} is a variant of \measure using $\operatorname{mean}$ instead of $\max$.
Formally, 
\begin{equation}\label{eq:baselinewomax}
\mathcal{T}(w, C; f)
=
\sum\nolimits_{\{v'_{1}, v'_{2}\} \in P(w)} \frac{ \frac{1}{\vert C \vert} \sum_{e \in C}f(w, e) \mathds{1}[v'_{1}, v'_{2} \in e]}
{\lvert P(w) \vert}.
\end{equation}

\smallsection{B8. \measure-non-$\mathbf{P(w)}$} \textbf{B8} is a variant of \measure without considering all the interactions in $P(w)$, but simply computes the maximum {value of} the group interaction function.
Formally, 
\begin{equation}\label{eq:baselineonlymax}
\mathcal{T}(w, C) = \max_{e \in C}(f(w, e)).
\end{equation}

\smallsection{B9. \measure-unnormalized} \textbf{B9} is a variant of \measure without normalizing the scores, but simply computes the summation of scores of all interactions in $P(w)$.
Formally,
\begin{equation}\label{eq:baselinenonorm}
    \mathcal{T}(w, C; f) =
    \sum\nolimits_{\{v'_{1}, v'_{2}\} \in P(w)} 
    \max_{e \in C} f(w, e) \mathds{1}[v'_{1}, v'_{2} \in e],
\end{equation}

\begin{figure}[t]
    \centering
    \vspace{-2mm}
    \includegraphics[width=0.35\textwidth]{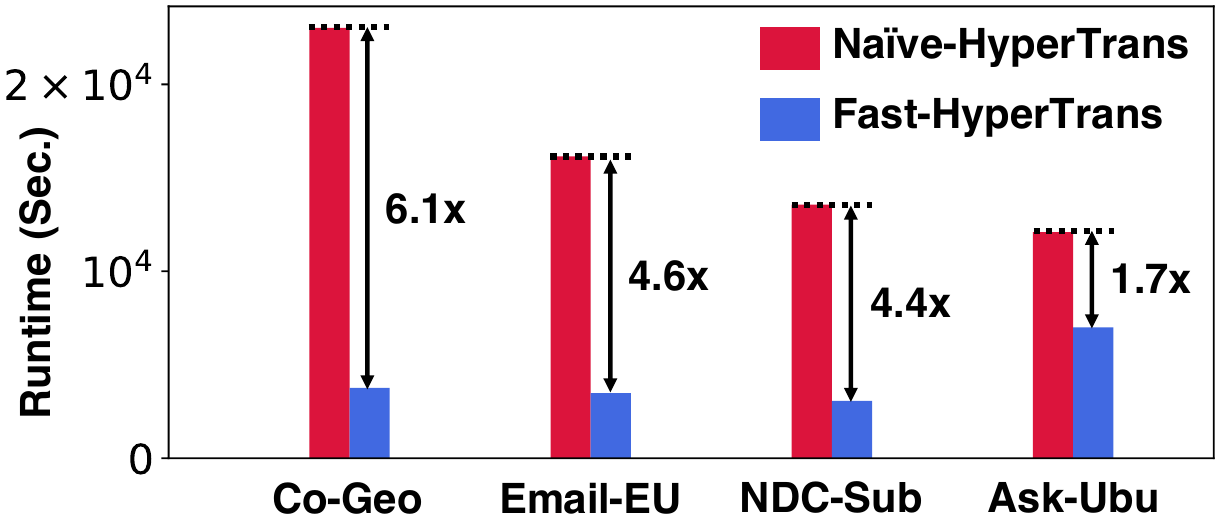}
    \vspace{-1mm}
    \caption{\label{fig:fasttime}Runtime of \naivemeasure and \fastmeasure (Algorithm~\ref{algo:fastmeasure}).}
    \vspace{-3mm}
\end{figure}

\begin{algorithm}[t!]
\small
\caption{\fastmeasure
\label{algo:fastmeasure}}
\KwIn{Hyperwedge $w$, candidate set $C$, and score function $f$.}
\KwOut{Hyperwedge transitivity $\mathcal{T}(w, C)$}
$\Phi(\{v'_{1}, v'_{2}\}) \leftarrow 0 , \forall \{ v'_{1}, v'_{2}\} \in P(w)$ \label{line:init_phi} \\
\ForEach{$e \in C$}  { \label{line:for_start}
    \ForEach{$\{v'_{1}, v'_{2}\} \in P(w) \cap \binom{e}{2}$} {\label{line:algo1pair1} 
        $\Phi(\{v'_{1}, v'_{2})\} \leftarrow \max(\Phi(\{v'_{1}, v'_{2}\}), f(w, e)) $\label{line:algo1replace} \\
    }
}
\Return $\sum_{\{v'_{1}, v'_{2}\} \in P(w)} \Phi(\{v'_1, v'_2\})  / |P(w)|$ \label{line:return_val}
\end{algorithm}
\vspace{-3mm}
\subsection{Fast \& Exact Computational Algorithm}
We propose a fast and exact algorithm \textbf{\fastmeasure} (Algorithm~\ref{algo:fastmeasure}) for computing \measure.
{In essence, \fastmeasure identifies $e\in C$ that maximizes $f(w, e) \mathds{1}[v'_{1}, v'_{2} \in e]$ in Eq~\eqref{eq:hypertrans}, without exhaustively considering all hyperedges in $C$, which results in reduced computation time.}
Specifically, given a target hyperwedge $w$, a candidate set $C$, and a score function $f$,
\fastmeasure first initializes the score of each interaction in $P(w)$ as $0$ (Line~\ref{line:init_phi}),
then for each candidate hyperedge $e \in C$, 
\fastmeasure records and updates the highest score for each interaction in $P(w)$ that is covered by $e$ (Lines~\ref{line:for_start}-\ref{line:algo1replace}).
Finally, \fastmeasure returns the average score as defined in Eq~\eqref{eq:hypertrans} (see Theorem~\ref{thm:exactness}).


\begin{theorem}[Exactness]\label{thm:exactness}
Given any $w, C, f$, \fastmeasure (Algorithm~\ref{algo:fastmeasure}) outputs $\mathcal{T}(w, C; f)$ as defined in Eq~(\ref{eq:hypertrans}).    
\end{theorem}
\vspace{-2mm}
\begin{proof}
See the online appendix~\cite{github}.
\end{proof}
We compare the efficiency of \fastmeasure with a naive computational method \naivemeasure, which computes \measure directly based on Eq~\eqref{eq:hypertrans}.
{That is, \naivemeasure exhaustively considers all hyperedges in $C$ to identify $e\in C$ that maximizes $f(w, e) \mathds{1}[v'_{1}, v'_{2} \in e]$ in Eq~\eqref{eq:hypertrans} (refer to the online appendix~\cite{github} for details).}
As a result, the time complexity of \fastmeasure is upper bounded by that of \naivemeasure, as formalized in Theorem~\ref{thm:fastcomplexity}.
\begin{theorem}[Time complexity]\label{thm:fastcomplexity}
    Given any $w, C, f$, $TC_{fast}(w,C) = \mathcal{O}(TC_{naive}(w,C))$,
    where $TC_{naive}(w,C,f)$ is the time complexity of \naivemeasure,
    and $TC_{fast}(w,C,f)$ is that of \fastmeasure.
\end{theorem}
\begin{proof}
    See the online appendix~\cite{github}.
\end{proof}

Theorem~\ref{thm:fastcomplexity} is supported by our experiments showing that \fastmeasure is consistently faster than \naivemeasure 
{for the computation of $T(G)$} on real-world hypergraphs (see Figure~\ref{fig:fasttime}).

%% file: 004Observations.tex

\begin{table}[t!]
    \vspace{-3mm}
	\caption{\label{tab:datastat} Descriptive statistics 
 (the number of nodes $\lvert V \vert$, the number of hyperedges $\lvert E \vert$, the number of hyperwedges $\lvert W \vert$, and the maximum hyperedge size $\max_{e\in E} \lvert e \vert$)
 of 12 real-world hypergraphs from 5 different domains.}
    \vspace{-1mm}
	\scalebox{0.8}{
		\begin{tabular}{l| r | r | r | r}
			\toprule
                \textbf{Data} &
                {\textbf{$\lvert V \vert $}} &
                {\textbf{$\lvert E \vert $}} &
                {\textbf{$\lvert W \vert $}} &
                {\textbf{$\max_{e\in E} \lvert e \vert$}} \\
			\midrule
			email-enron & 143 & 1,459 & 80,715 & 37 \\
                email-eu & 986 & 24,520 & 8,392,205 & 40 \\
                \midrule
                ndc-classes & 1,149 & 1,049 & 32,005 & 39 \\
                ndc-substances & 3,767 & 6,631 & 2,347,653 & 187 \\
                \midrule
                contact-high & 242 & 12,704 & 585,246 & 5 \\
                contact-primary & 327 & 7,818 & 2,221,968 & 5 \\
                \midrule
                coauth-dblp & 1,836,596 & 2,170,260 & 121,513,272 & 280 \\
                coauth-geology & 1,091,979 & 909,325 & 36,564,161 & 284 \\
                coauth-history & 503,868 & 252,706 & 1,536,732 & 925 \\
                \midrule
                qna-ubuntu & 90,054 & 115,987 & 21,526,221 & 14 \\
                qna-server & 152,658 & 222,610 & 94,719,715 & 66 \\
                qna-math & 33,541 & 86,730 & 27,648,084 & 209 \\
			\bottomrule
		\end{tabular}}
\vspace{-0.5mm}
\end{table}

In this section, we examine the transitivity patterns in real-world hypergraphs using the \measure measure.
We observe and demonstrate that the transitivity patterns in real-world hypergraphs differ significantly from those in null hypergraphs.
{Throughout the section, we use all hyperedges as the candidate set, i.e., $C=E$.}

\subsection{Datasets} 
We use 12 real-world hypergraphs from 5 different domains, 
after removing duplicated hyperedges and self-loops.
The descriptive statistics of the datasets are in Table~\ref{tab:datastat}.
The \textit{email}, \textit{drug}, \textit{contact}, \textit{coauthorship} datasets, and \textit{qna-ubuntu} dataset are from Benson et al.~\cite{benson2018simplicial}, while the other \textit{qna} datasets are from Kim et al.~\cite{kim2022reciprocity}.
\begin{enumerate}[wide, labelindent=5pt]
    \item \textbf{email}: each node represents a user, and each hyperedge represents an email, containing the email's sender, receivers, and CCs.
    \item \textbf{drug}: each node represents a class (substance), and each hyperedge represents a drug, containing the drug's classes (substances)
    \item \textbf{contact}: each node represents a person, and each hyperedge represents an instance of group communication, containing the people participating in the communication.
    \item \textbf{coauthorship}: each node represents a researcher, and each hyperedge represents a publication, containing the coauthors.
    \item \textbf{qna}: each node represents a user, and each hyperedge represents a question, containing the users asking or answering it.
\end{enumerate}


\begingroup
\begin{table}[t]
\vspace{-3mm}
\caption{(Observation~\ref{obs:hypergraph}) Hypergraph transitivity.
Real-world hypergraphs are usually more transitive than their random counterparts, while the \textit{qna} datasets show the opposite trend.
All statistics are significant under $\alpha = 0.05$. * indicates hypergraph transitivity $< 10^{-3}$ and ** indicates P-value $< 10^{-2}$.} 
\label{tab:observation1}
\vspace{-1mm}
    \centering
    \scalebox{0.8}{
    \renewcommand{\arraystretch}{0.9}
        \centering
        \begin{tabular}{l| c | c | c | c }
            \toprule
            \textbf{Data} & \textbf{Real} & \textbf{HyperCL} & \textbf{Z-stat} & \textbf{P-value}\\
            \midrule
            \midrule
            email-enron & 0.195 & 0.078 & 378.3& 0.00**\\
            email-eu &0.125 & 0.053 & 240.1& 0.00**\\
            \midrule
            ndc-classes & 0.052 & 0.008 &146.7& 0.00**\\
            ndc-substances & 0.019 & 0.005 &47.3& 0.00**\\
            \midrule
            contact-high & 0.345 & 0.119 &764.7& 0.00**\\
            contact-primary & 0.336 & 0.223 &380.7& 0.00**\\
            \midrule
            coauth-dblp & 0.007 & 0.000* &23.2& 0.00**\\
            coauth-geology & 0.005 & 0.000* &16.6& 0.00**\\
            coauth-history & 0.002 & 0.000* &6.6& 0.00**\\
            \midrule
            qna-ubuntu & 0.005 & 0.014 &32.0 & 0.00**\\
            qna-server & 0.005 & 0.017 &38.3& 0.00**\\
            qna-math & 0.025 & 0.040 &46.6& 0.00**\\
            \bottomrule
        \end{tabular}
        }
\end{table}
\endgroup
\begingroup
\begin{table}[t!]
    \vspace{-3mm} 
	\caption{\label{tab:obsbodyandtrans} (Observation~\ref{obs:hyperwedge}) 
            Spearman's rank correlation coefficient between the body group sizes and the hyperwedge transitivities. 
            Positive correlations between them consistently exist in real-world hypergraphs, which become weaker or even opposite in their random counterparts.
            The positive correlation cell is in \textcolor{dodgerblue}{blue}, the negative cell is in \textcolor{backredcolor}{red}, and 0.00* indicates a correlation coefficient between -0.01 and 0.01.}
        \vspace{-1mm}
	\scalebox{0.8}{
        \renewcommand{\arraystretch}{0.9}{
        \centering
            \begin{tabular}{l| c | c | c }
			\toprule
                \textbf{Data} &\textbf{Real} & \textbf{HyperCL} & \textbf{\generator} \\
			\midrule
                \midrule
			email-enron & \posc 0.09 & \negc -0.09 & \posc 0.23 \\
                email-eu & \posc 0.12 & \negc -0.14 & \posc 0.22 \\
                \midrule
                ndc-classes & \posc 0.32 & \negc -0.10 & \posc 0.40 \\
                ndc-substances & \posc 0.14 & \negc -0.10 & \posc 0.24 \\
                \midrule
                contact-high & \posc 0.13 & 0.00* & \posc 0.29  \\
                contact-primary & \posc 0.13 & 0.00* & \posc 0.30 \\
                \midrule
                coauth-dblp & \posc 0.12 & 0.00* & \posc 0.20  \\
                coauth-geology & \posc 0.14 &  0.00* & \posc 0.26 \\
                coauth-history & \posc 0.12 & \posc 0.05 & \posc 0.19 \\
                \midrule
                qna-ubuntu & \posc 0.04 & 0.00* & \posc 0.03 \\
                qna-server & \posc 0.04 & 0.00* & \posc 0.04 \\
                qna-math & \posc 0.04  & \posc 0.01 & \posc 0.13  \\
			\bottomrule   
		\end{tabular}        
            }
	}    
\end{table}
\endgroup

\subsection{Observations}\label{subsec:mainobservation}
We investigate the transitivity patterns in real-world hypergraphs at four different levels: hypergraphs, hyperwedges, nodes, and hyperedges.
We use \textbf{HyperCL}~\cite{lee2021hyperedges} as a null hypergraph model, which preserves the expected degree distribution of real-world hypergraphs, to generate the random counterpart of each real-world dataset using its statistics (spec., degree distribution and hyperedge size distribution).
We shall show that the patterns in real-world hypergraphs differ significantly from those in random ones.

\smallsection{L1: hypergraph level}
We compute the hypergraph transitivity $T$ of the real-world and random hypergraphs.
{As shown in Table~\ref{tab:observation1}, 
the real-world hypergraphs are more transitive than their random counterparts on all datasets except for the \textit{qna} datasets.}
All the numerical comparisons between the real-world and random hypergraphs are statistically significant at a significance level of $\alpha = 0.05$
(see {Table~\ref{tab:observation1}} and Appendix~\ref{subsec:expresult} for  details).
\vspace{-1mm}
\begin{observation}\label{obs:hypergraph}
Real-world hypergraphs are usually more transitive than their random counterparts.
However, the \textit{qna} datasets show the opposite tendency.
\end{observation}
\vspace{-1mm}

\smallsection{L2: hyperwedge level} 
At the hyperwedge level, we investigate for each hyperwedge the relationship between the size of its body group and its transitivity.
For each dataset, we measure the Spearman's rank correlation coefficient~\cite{fieller1957tests} between the sequence of body group sizes $\lvert B(w)\vert$'s and 
that of the hyperwedge transitivity values $\mathcal{T}(w)$'s.
As shown in Table~\ref{tab:obsbodyandtrans}, consistently positive correlations are observed on the real-world hypergraphs,
while such correlations become very weak or even negative on their random counterparts.
Intuitively, this implies that in real-world scenarios,
groups sharing many \textit{`common friends'} are more likely to interact.

\begin{observation}\label{obs:hyperwedge}
Consistently positive correlations exist between the body group sizes and hyperwedge transitivities in real-world hypergraphs, which cannot be observed in their random counterparts.
\end{observation}

\begin{figure}[t]
    \centering
    \vspace{-3mm}
    \includegraphics[width=0.45\textwidth]{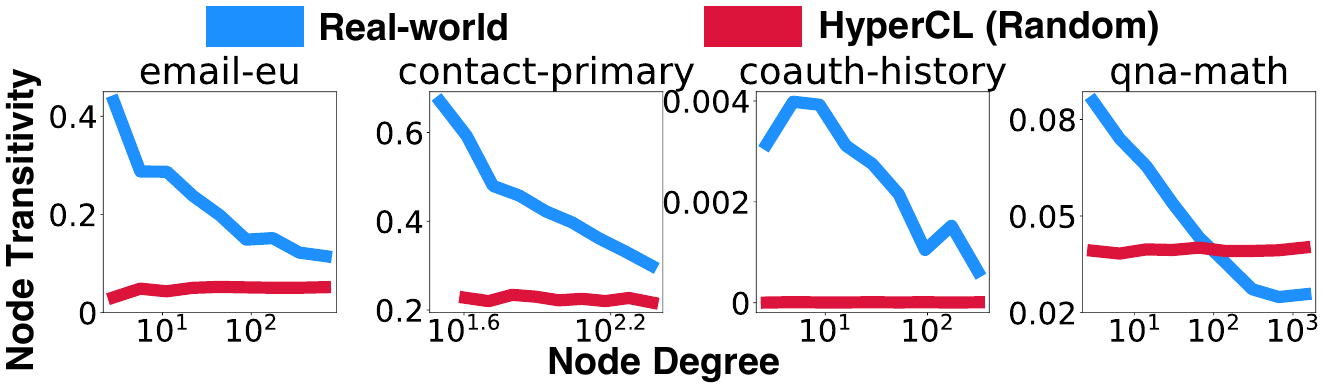}   
    \vspace{-1mm}
    \caption{\label{fig:obsnode}(Observation~\ref{obs:node}) Relation between the degree of a node and its transitivity in the \textcolor{dodgerblue}{real-world}  and \textcolor{crimson}{random} hypergraphs.    As the degree of a node increases, its transitivity tends to decrease in real-world hypergraphs, while such patterns are not observed in random counterparts.}
    \vspace{-1mm}
\end{figure}


\begin{table}[t]
    \centering\caption{\label{tab:obshyperedge}(Observation~\ref{obs:hyperedge}) The range of hyperedge transitivity.
In real-world hypergraphs, the ranges are much wider than in their random counterparts generated by HyperCL, while \generator reproduces ranges similar to the real-world ones.} 
    \scalebox{0.8}{
    \renewcommand{\arraystretch}{0.9}
        \centering
        \begin{tabular}{l| c | c | c}
            \toprule
            \textbf{Data} & \textbf{Real} & \textbf{HyperCL} & \textbf{\generator} \\
            \midrule
            \midrule
            email-enron & 0.725 & 0.279 & 0.732\\
            email-eu &0.809 & 0.248 & 0.792 \\
            \midrule
            ndc-classes & 0.600 & 0.075 & 0.410 \\
            ndc-substances & 1.000 & 0.032 & 0.411 \\
            \midrule
            contact-high & 0.794 & 0.316 & 0.768 \\
            contact-primary & 0.693 & 0.395 & 0.839 \\
            \midrule
            coauth-dblp & 1.000 & 0.105 & 1.000 \\
            coauth-geology & 1.000 & 0.069  & 1.000 \\
            coauth-history & 1.000 & 0.333 & 1.000 \\
            \midrule
            qna-ubuntu & 0.667 & 0.500 & 1.000 \\
            qna-server & 0.667 & 0.333 & 1.000 \\
            qna-math & 0.667 & 1.000 & 1.000 \\
            \bottomrule
        \end{tabular}
        }
\end{table}

\smallsection{L3: node level} 
We investigate for each node $v$, the relationship between its degree and the transitivities of the hyperwedges \textit{`around'} $v$.
The degree of a node $v$ is $d(v) = \lvert \{e \in E : v \in e \}\vert$,
and the set of the hyperwedges \textit{`around'} $v$ is $W_{v} = \{w \in W : v \in B(w)\}$ consisting of
those including $v$ in their body group.
We define the transitivity of each node $v$ as 
$\mathcal{T}(v) = \frac{1}{\lvert W_{v} \vert }\sum_{w \in W_{v}} \mathcal{T}(w)$.
Figure~\ref{fig:obsnode} illustrates the trend between node degrees and transitivities,
where we process the data points by logarithmic binning w.r.t degrees.
{On the real-world hypergraphs, 
the average transitivity of nodes decreases as the node degree increase,
while such trends cannot be observed on their random counterparts.}
This observation is in line with the previous results on graphs~\cite{ravasz2003hierarchical, zhou2005maximal},
where the transitivity of a node $v$ is often negatively correlated to its degree.
Results on the other datasets are in the online appendix~\cite{github}.

\begin{observation}\label{obs:node}
In real-world hypergraphs, the transitivity of a node is negatively correlated to its degree, which cannot be observed in their random counterparts.
\end{observation}

\smallsection{L4: hyperedge-level}
For each hyperedge $e$, let $W_{e}$ denote the set of hyperwedges including $e$ (i.e., $W_{e} = \{w \in W : e \in w\}$).
We define the transitivity of each hyperedge $e$ as $\mathcal{T}(e) = \frac{1}{\lvert W_{e}\vert }\sum_{w \in W_{e}} \mathcal{T}(w)$.
We further define the \textit{\textbf{range}} of hyperedge transitivity (of a hypergraph) as $\max_{e \in E}\mathcal{T}(e) - \min_{e \in E}\mathcal{T}(e)$.
As reported in Table~\ref{obs:hyperedge}, the range of hyperedge transitivity of the real-world hypergraphs is wider than that of random counterparts.

\begin{observation}\label{obs:hyperedge}
Real-world hypergraphs have significantly wider ranges of hyperedge transitivity than their random counterparts.
\end{observation}

%% file: 005Generator.tex
\begin{figure}[t]
    \centering
    \vspace{-1mm}
    \subfigure[Hypergraph transitivity v.s community size.]
    {\includegraphics[width=0.22\textwidth]{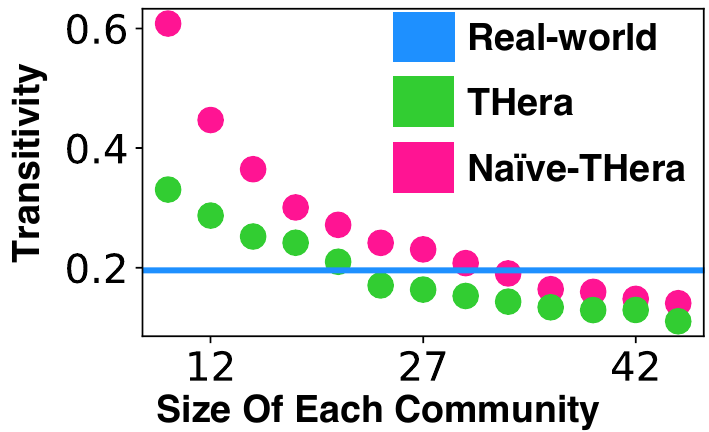}}
    \hspace{3mm}
    \subfigure[The cumulative distribution function (CDF) of node degrees.]
    {\includegraphics[width=0.22\textwidth]{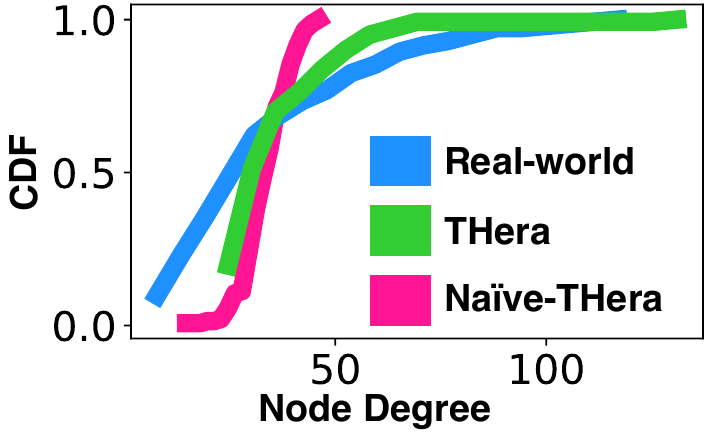}}
    \vspace{-1mm}
    \caption{\label{fig:intracommunity} \generator can control the transitivity of the generated hypergraph and it fits the real-world node-degree distribution better than its naive version \naivegenerator.}
\end{figure}
\input{main_alg.tex}

\input{main_res.tex}

We have observed that real-world hypergraphs have different transitivity patterns from their random counterparts generated by HyperCL.
In this section, we introduce a scalable hypergraph generator \textbf{\generator} (\myuline{\textbf{T}}ransitive \myuline{\textbf{H}}ypergraph gen\underline{\textbf{ERA}}tor), which reproduces the observed real-world transitivity patterns.


\subsection{Proposed Generator: \generator}\label{subsec:generator}

\smallsection{High-level ideas} {According to Observation~\ref{obs:hypergraph}, a realistic hypergraph generator should produce hypergraphs with notably higher transitivity when compared to the null hypergraph model (HyperCL). To enhance transitivity, we leverage the understanding provided by \textsc{Axiom}~\ref{ax:incre_changes_replacement}, which indicates that hyperwedges exhibiting extensive overlap with other hyperedges tend to have higher transitivity than those with less overlap. In our approach, we utilize the community structure of nodes to encourage hyperedges to overlap with one another, thereby promoting increased transitivity.}

{Our preliminary approach, \naivegenerator, assigns each node  to a community, and creates intra-community hyperedges among nodes sampled uniformly at random within each community.\footnote{The size of each hyperedge is sampled from the ground-truth hyperedge size distribution. All community has the same size, which is a hyperparameter.} By using \naivegenerator, we can control the hypergraph transitivity value by adjusting the community sizes (see Figure~\ref{fig:intracommunity}(a)), resulting in, however, hypergraphs with uniform divisions and near-uniform degree distributions, which are unrealistic \cite{ko2022growth,do2020structural} (see Figure~\ref{fig:intracommunity}(b)).}


{Our proposed generator, \generator, addresses these limitations of \naivegenerator by introducing {inter-community} hyperedges and producing realistic degree distributions. To achieve this, \generator assumes a hierarchical structure of nodes, represented as a tree, and assigns each node to a level in the tree.
The nodes at each level are split into disjoint communities.
Then, \generator generates two types of hyperedges consisting of  (1) nodes sampled \textit{``locally''} within the same community or
(2) nodes sampled \textit{``globally''} among all existing ones.
In global hyperedges, which connect different communities, there is a higher chance of selecting nodes from lower levels, leading to realistic skewed degree distributions (see Figure~\ref{fig:intracommunity}(b)).}

\smallsection{Algorithmic details}
In Algorithm~\ref{algo:generator}, we provide pseudocode of \generator, which introduces new nodes and generates new hyperedges in an incremental way.
The inputs of \generator are: 
\vspace{-1mm}
\begin{enumerate}[wide, labelindent=1pt]
    \item the number of nodes $n$ and the hyperedge size distribution $S$;\footnote{For each $k$, $S(k) \in \mathds{N}$ denotes the expected number of hyperedges of size $k$,
    which is the ground-truth value from the real-world hypergraphs in our experiments.\label{foot:defofm}}
    \item the community size (i.e., the number of nodes in each community) $C$, and the ratio of intra-community hyperedges $p$;
    \item $\alpha \in [1, \infty]$, which controls the likelihood of nodes at different levels being included in the generated hyperedges, and $\beta \in \mathds{N}^{+}$, which controls the number of nodes at each level.
\end{enumerate}
\vspace{-1mm}
The output of \generator is a hypergraph having $n$ nodes and $m = \sum_k S(k)$ hyperedges with an expected hyperedge size distribution equal to $S$.
\generator first distributes $m$ hyperedges to the $n$ nodes (Lines~\ref{line:nedgestart}-\ref{line:nedgeend}) so that $AE(i)$ hyperedges are {newly} generated for each node $v_i$.
After, \generator puts a single node $v_1$ at level $0$, 
then starting from the level $T = 1$, at each level,
\generator generates $T^{\beta}$ communities, each of which contains $C$ nodes (Line~\ref{line:currentlevel}).
For each node $v_{idx}$ at level $T$, \generator generates $AE(idx)$ hyperedges {that contain $v_{idx}$ and follow} the size distribution $S$.
Each hyperedge becomes either an intra-community one (see \textbf{\texttt{IntraCommunity}}\textbf{\texttt{Generate}}), with the probability of $p$, 
{or a global one (see \textbf{\texttt{Hierarchical}}\textbf{\texttt{Generate}}), with the probability of $1-p$.}
For $v_{idx}$, \textbf{\texttt{IntraCommunity}}\textbf{\texttt{Generate}} samples nodes within the community where $v_{idx}$ belongs to, while
\textbf{\texttt{Hierarchical}}\textbf{\texttt{Generate}} samples nodes from the nodes at the level equal to or lower than the current level,
where the probability of a node at level $\ell$ being sampled is proportional to $\alpha^{-\ell}$.
This ensures that nodes at lower levels are more likely to be included when $\alpha \geq 1$ (see {Proposition}~\ref{prop:high_level_low_degree}).

\begin{proposition}[Negative correlation between layer index and node degree]\label{prop:high_level_low_degree}
    {For any $v_1, v_2$ with $L(v_1) > L(v_2)$ and $\alpha \geq 1$, in a hypergraph generated by \generator (Algorithm~\ref{algo:generator}), the expected degree of $v_1$ is smaller than that of $v_2$, i.e., $\mathds{E}[d(v_1)] < \mathds{E}[d(v_2)]$.}
\end{proposition}
\vspace{-3mm}
\begin{proof}
    See the online appendix~\cite{github}.
\end{proof}
\vspace{-3mm}


\subsection{Empirical Evaluation of \generator}\label{subsec:generatoreval}
{We conduct a comparative analysis to assess the ability of \generator to replicate observed real-world transitivity patterns.
As competitors, we consider \textbf{HyperPA} \cite{do2020structural}, \textbf{HyperFF} \cite{ko2022growth}, \textbf{HyperLap} \cite{lee2021hyperedges}, and \textbf{HyperLap+} \cite{lee2021hyperedges}, which all aim to create realistic hypergraphs. 
It is worth noting that our proposed generator, along with \textbf{HyperPA} and \textbf{HyperFF}, adds nodes and hyperedges incrementally, offering two advantages: (1) modeling the evolution of hypergraphs, and (2) serving as benchmarks for temporal hypergraph algorithms. See Appendix~\ref{subsec:expsetting} for  hyperparameter settings.
However, HyperLap(+)~\cite{lee2021hyperedges} does not provide these advantages.\footnote{It also requires a realistic degree distribution as an input, while \generator does not.}}
{Given an input real-world hypergraph, each generator approximates it by generating a hypergraph of a similar scale, using its statistics.}

\smallsection{Reproduction of Observation~\ref{obs:hypergraph}}
We measure the transitivity of hypergraphs generated by \generator, four baseline methods, and the null model {(i.e., HyperCL).}
Among all the six methods, \generator generates hypergraphs with transitivity values closest to those of the real-world hypergraphs (see Table~\ref{tab:observation1}). 
In addition, \generator preserves the hyperwedge transitivity distribution most accurately among the six methods. 
We numerically measure the Kolmogorov–Smirnov D-Statistics~\cite{fieller1957tests} between the hyperwedge transitivity distribution of real-world hypergraphs and that of generated hypergraphs. 
As shown in Table~\ref{tab:genobservation1}, the overall ranking over the entire datasets of \generator is the highest among all the methods.

\smallsection{Reproduction of Observation~\ref{obs:hyperwedge}}
We have verified that there is a positive correlation between the body group sizes and the transitivity values of hyperwedges in the real-world hypergraphs (see Observation~\ref{obs:hyperwedge} and Table~\ref{tab:obsbodyandtrans}).
We now investigate whether this tendency also exists in the hypergraphs generated by \generator. 
As demonstrated in the last column of Table~\ref{tab:obsbodyandtrans}, \generator successfully reproduces this pattern, exhibiting a positive correlation between the body group sizes and the transitivity values of hyperwedges.

\begin{figure}[t!]
    \vspace{-2mm}
    \centering
    \includegraphics[width=0.45\textwidth]{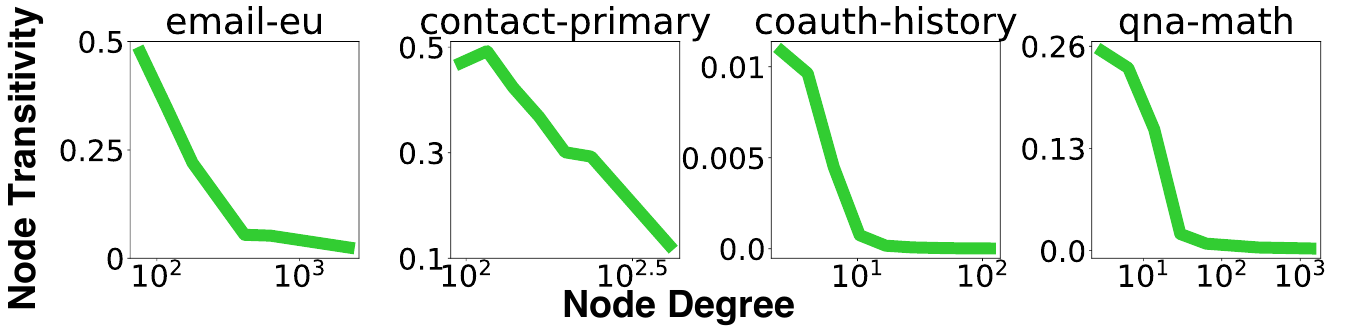}
    \vspace{-1mm}
    \caption{\label{fig:gennode}The relationship between {the degree and transitivity of nodes} in the hypergraphs generated by \generator.}
\end{figure}

\smallsection{Reproduction of Observation~\ref{obs:node}} 
At the node level, {hyperwedges} \textit{``around"} a high-degree node tend to have low transitivity (see observation~\ref{obs:node} and Figure~\ref{fig:obsnode}).
We now examine whether such a pattern is also present in the hypergraphs generated by \generator. 
As depicted in Figure~\ref{fig:gennode}, using the same plotting method described in observation~\ref{obs:node}, the decreasing trend of the \textcolor{limegreen}{green} lines shows similarities with the real-world scenarios.
Therefore, we conclude that \generator reproduces the real-world transitivity pattern at the node level.

\smallsection{Reproduction of Observation~\ref{obs:hyperedge}}
The range of hyperedge transitivity is much broader in the real-world hypergraphs than in the random ones generated by the null model (i.e. {HyperCL}).
We now investigate the range of hyperedge transitivity in the hypergraphs generated by \generator. 
As shown in Table~\ref{tab:obshyperedge}, the ranges generated by \generator are much closer to the real-world ones, compared to those generated by the null model.
Notably, in some datasets, the range generated by \generator matches the actual range.

\smallsection{Reproduction of other properties} 
{We compare the ability of \generator and the baseline generators to produce realistic hypergraphs with respect to density, diameter, hyperedge size, degree, and intersection size.
In summary, \generator outperforms the other generators with respect to density and achieves competitive results with respect to other measures.  For details, refer to Appendix~\ref{subsec:expresult}.}
\subsection{Scalability of \generator}\label{subsec:generatorscalability}
In this subsection, we analyze the scalability of \generator.
{We first examine the time complexity of generating a hypergraph with a specified node count and hyperedge sizes.}
\textbf{\texttt{CommunityGeneration}} takes $\mathcal{O}(1)$ to sample nodes within a community, since it is equivalent to uniform sampling~\cite{schwarz2011darts}.
In \textbf{\texttt{HierarchicalGeneration}}, the level $\ell$ can be chosen in $\mathcal{O}(\log T) = \mathcal{O}(\log |V|)$ time~\cite{bringmann2012efficient},
and {a} node can be further chosen in $\mathcal{O}(1)$ time by a uniform sampling within a level.
In sum, the time complexity to generate a hyperedge of size $\lvert e \vert $ is $\mathcal{O}(\lvert e\vert \log_{2}{\vert V \vert})$,
and thus the total time complexity to generate all the hyperedges is $\mathcal{O}(\log_{2}{\vert V \vert}  \sum_{e \in E}\lvert e\vert)$.
For empirical verification, we measure the runtime of \generator when it generates synthetic hypergraphs.
The synthetic hypergraphs are obtained by scaling up the \textit{email-enron} dataset by $10^{2}$ to $10^{4.5}$ times.
As shown in Figure~\ref{fig:genlinear}, the runtime of \generator is linear,
and \generator can generate a hypergraph with $10^{7.5}$ hyperedges within a minute.
Moreover, the generation process of \generator is terminated within {a few} seconds for all the real-world hypergraphs, where the largest dataset \textit{coauth-dblp} has more than $2.1$ million hyperedges (see Table~\ref{tab:genruntime}).

Regarding memory, on top of the memory needed to save the generated hypergraph, which is a common cost for each generator,
\generator only requires a hashtable $\Psi_L$ of size $\mathcal{O}(\lvert V \vert)$,
a list $AE$ of size $\mathcal{O}(\lvert V \vert)$,
and a hyperedge size distribution $S$ of size $\mathcal{O}(\lvert E \vert)$,
which gives $\mathcal{O}(\lvert V \vert + \lvert E \vert)$ total additional memory requirement.

{We also compare the empirical speed and memory consumption of \generator and the four baseline generators for approximating the six largest datasets, where $\vert V \vert > 10^{4}$. \generator exhibits the shortest runtime for all six datasets, and it exhibits the lowest memory consumption for four of the datasets.
See Appendix~\ref{subsec:appgenscale} for details.}

%% file: main_alg.tex
\begin{algorithm}[t]
\small
\caption{\generator: Transitive hypergraph generator\label{algo:generator}}
\KwIn{(1) Number of nodes $n$, hyperedge size distribution $S$ \newline 
(2) Community size $C$, intra-community hyperedge ratio $p$ \newline
(3) Level-sampling coefficient $\alpha$, level-size coefficient $\beta$}
\KwOut{Generated hypergraph $G' = (V', E')$}

$idx \leftarrow 1$; $T \leftarrow 0$; $\Psi_L(T') \leftarrow \emptyset, \forall T' \in \mathds{N}^+$; $E' \leftarrow \emptyset$; $m \leftarrow \sum_k S(k)$ \\
$AE(1) \leftarrow 0$; $AE(i) \leftarrow 1, \forall i = \{2, \cdots, n\}$; $\Psi_{L}(0) \leftarrow \{v_{1}\} $ \label{line:nedgestart}\\ 
\While {$\operatorname{sum}(AE) < m$}{
$a \sim \operatorname{discrete-uniform}(\{2 , \cdots n\})$; $AE(a) \leftarrow AE(a) + 1$\label{line:nedgeend}\\ 
}

\While{$idx < n$}{   
    $T \leftarrow T + 1 $\Comment{{level in the hierarchy}}  \\
    $\Psi_{L}(T) \leftarrow\{v_{(idx + i)}\}_{1 \leq i \leq \min(CT^{\beta}, n - idx)}$ \label{line:currentlevel}
    \Comment{{node set at level $T$}} 
    \\
    \For {$i = 1 \ \textbf{to} \  \min (CT^{\beta}, n - idx) $}{
        $idx \leftarrow idx + 1$ \\ 
        \For {$j = 1 \ \textbf{to} \ AE(idx)$ }{
            $e' \leftarrow \{v_{idx}\}$; $s \sim S$; $q \sim \operatorname{uniform}(0, 1)$ \label{line:sample_prep} \\
            \If {$q$ < $p$} {
                $e'{\leftarrow}$ \textbf{\texttt{IntraCommunityGenerate}}($e'{,}idx{,}C{,}T{,}s{,}\Psi_{L}$) \\
                }
            \If {$\lvert e'\vert$ < s} {
                $e' \leftarrow$ \textbf{\texttt{HierarchicalGenerate}}($e', T, s, \Psi_{L}$)
                }
            $E' \leftarrow E' \cup \{ e' \}$
            }
        }
    }
\textbf{return} $G' = (V' = \{v_{1} , \cdots , v_{n}\}, E')$  \\ 
 \nlnonumber
\nonumber \hrulefill \\
\setcounter{AlgoLine}{0}
\nlnonumber
\SetKwProg{myproc}{\texttt{IntraCommunityGenerate}}{($e', {idx}, C, T, s, \Psi_L$)}{}~\label{subalgo:intracom}
  \myproc{}{
  ${idx'} = C \times \ceil[\big]{({idx}-2)/C} + 1$ \Comment{beginning index of community} \label{line:intracombeg}\\ 
  $V_{C} \leftarrow \Psi_L(T) \cap \{v_{({idx'} + 1)}, \cdots ,v_{({idx'} + C)}\} \setminus e'$ \label{line:intracomend}\Comment{community}\\
  $V' \leftarrow $ uniformly sample $\min(s - 1, |V_C|)$ nodes from $V_{C}$ \\ 
  \Return $e' \cup V'$ \\
  }  
\setcounter{AlgoLine}{0}
\nlnonumber
\SetKwProg{myproce}{\texttt{HierarchicalGenerate}}{($e', T, s, \Psi_{L})$}{}~\label{subalgo:hier}
  \myproce{}{ 
  \While{$\lvert e'\vert < s$}{
    $\ell \leftarrow $ sample a level from $[0, 1, \cdots , T] $ proportional to $[\lvert \Psi_{L}(0)\vert, \alpha^{-1}\lvert \Psi_{L}(1)\vert , \cdots , \alpha^{-T}\lvert \Psi_{L}(T)\vert]$ respectively \label{line:firstsample}\\
    $v' \leftarrow $ sample a node from $\Psi_{L}(\ell)$ uniformly at random \label{line:secondsample}\\ 
    $e' \leftarrow e' \cup \{v'\}$
    }
    \Return $e'$ \\
  }  
\end{algorithm}

%% file: main_res.tex
\begin{table*}[t!]
\vspace{-2mm}
\caption{Hypergraph transitivity $T(G)$ and D-statistic {between} hyperwedge transitivity $\mathcal{T}(w)$ distributions {in} the real-world hypergraphs and the generated ones.
Among six generators, \generator reproduces $T(G)$ and distribution of $\mathcal{T}(w)$ most accurately.
The best reproduction results on each dataset are colored, and `*' indicates that the value is less than $10^{-3}$.
{`-' indicates that the generation process either exceeds the time limit of 12 hours or encounters an out-of-memory issue.}
} \label{tab:genobservation1}
\vspace{-1mm}
    \centering
    \scalebox{0.8}{
    \renewcommand{\arraystretch}{0.9}
        \centering
        \begin{tabular}{c | c| c c |c c |c c| c c c| c c c | c}
            \toprule
            \multirow{2}{*}{\textbf{Statistic}} & \multirow{2}{*}{\textbf{Generator}} &  \multicolumn{2}{c|}{email} & \multicolumn{2}{c|}{NDC} & \multicolumn{2}{c|}{contact} & \multicolumn{3}{c|}{coauthorship} & \multicolumn{3}{c|}{q\&a} & Average \\
            & & enron & eu & classes & substances & high & primary & dblp & geology & history & ubuntu & server & math  & ranking \\
            \midrule
            \midrule
            \multirow{7}{*}{\makecell{ \textbf{Hypergraph} \\ \textbf{transitivity} \\ \textbf{$T(G)$}}}&\textbf{Real World} & 0.195 & 0.125 & 0.052 & 0.019 & 0.345 & 0.336 & 0.007 & 0.005 & 0.002 & 0.005 & 0.005 & 0.025 & Real\\
            \cmidrule{2-15}
            
            &\textbf{\generator} & \goat 0.192 &  0.124 & \goat 0.052 & \goat 0.019 & \goat 0.344 & \goat 0.334 & \goat 0.007 & \goat 0.005 & \goat 0.002 & \goat 0.004 & \goat 0.004 & \goat 0.025  & \goat 1.08\\
            \cmidrule{2-15}
            
            &\textbf{HyperCL}~\cite{lee2021hyperedges} & 0.078 & 0.053 & 0.008 & 0.005 & 0.119 & 0.223 & 0.000* & 0.000* & 0.000* & 0.014 & 0.017 & 0.040  &  4.08\\
            
            &\textbf{HyperPA}~\cite{do2020structural} & 0.090 & 0.110 & 0.070 & - & 0.121 & 0.153 & - & - & - & 0.003 & - & - & 4.75\\
            
            &\textbf{HyperFF}~\cite{ko2022growth} & 0.176 & \goat 0.125 & 0.006 & 0.003 & 0.006 & 0.007 & 0.047 & 0.048 & 0.048 & 0.051 & 0.050 & 0.054 &  4.83\\
            
            &\textbf{HyperLap}~\cite{lee2021hyperedges} & 0.123 & 0.085 & 0.008 & 0.008 & 0.220 & 0.301 & 0.001 & 0.000* & 0.000* & 0.016 & 0.015 & 0.004 & 3.25\\

            &\textbf{HyperLap+}~\cite{lee2021hyperedges} & 0.231 & 0.144 & 0.026 & 0.016 & 0.322 & 0.338 & 0.042 & 0.019 & 0.005 & 0.029 & 0.023 & 0.007 & 3.54\\
            \midrule 
            \midrule 
            
            \multirow{6}{*}{\makecell{\textbf{D-Statistic from} \\ \textbf{real-world $\mathcal{T}(w)$} \\ 
            \textbf{distribution}}} &\textbf{\generator} & 0.137 & 0.186 & 0.208 & 0.187 & \goattwo 0.101 & \goattwo 0.099 & \goattwo 0.111 & 0.197 & \goattwo 0.066 & 0.035 & 0.170 & 0.101 & \goattwo 2.25\\
            \cmidrule{2-15}
            &\textbf{HyperCL} & 0.285 & 0.285 & 0.395 & 0.482 & 0.372 & 0.243 & 0.239 & 0.385 & 0.113 & \goattwo 0.020 & 0.054 & 0.084 & 3.75 \\
            &\textbf{HyperPA}~\cite{do2020structural} & 0.235 & 0.319 & 0.410 & - & 0.360 & 0.319 & - & - & - & 0.034 & - & - & 5.25\\
            
            &\textbf{HyperFF}~\cite{ko2022growth} & \goattwo 0.094 & \goattwo 0.158 & 0.638 & 0.831 & 0.631 & 0.699 & 0.126 & \goattwo 0.131 & 0.290 & 0.354 & 0.215 & 0.087 & 3.75\\
            
            &\textbf{HyperLap}~\cite{lee2021hyperedges} & 0.191 & 0.178 & 0.369 & 0.326 & 0.223 & 0.124 & 0.150 & 0.265 & 0.089 & 0.021 & \goattwo 0.051 & \goattwo 0.051 & 2.50\\
            
            &\textbf{HyperLap+}~\cite{lee2021hyperedges} & 0.262 & 0.244 & \goattwo 0.175 & \goattwo 0.185 & 0.166 & 0.180 & 0.509 & 0.405 & 0.119 & 0.061 & 0.104 & 0.184 & 3.23\\
            
            \bottomrule
        \end{tabular}
        }
\end{table*}

%% file: 006RelatedWork.tex
\smallsection{Transitivity in graphs}
Transitivity in graphs measures 
the likelihood of an edge existing between two neighbors of a node. 
Broadly speaking, there are local transitivity measure~\cite{watts1998collective} and global transitivity measure~\cite{newman2001random} in graphs.
Local transitivity is defined as
$\mathcal{T}(v) = \lvert \{ \{v_{i} ,v_{j}\} \in \binom{\mathcal{N}(v)}{2} : \{v_{i} ,v_{j}\} \in E \}\vert / \lvert {\mathcal{N}(v) \choose 2}\vert $
where $\mathcal{N}(v)$ is a neighborhood of a node $v$. 
Global transitivity {quantifies} the overall likelihood of the connection between node pairs sharing common neighbors in the whole graph,
which is defined as
$\mathcal{T}(G) = 3 \times \Delta / \lvert W(G) \vert $,
where $\Delta$ is the number of triangles.
Alternatively, one can average the local transitivity values (i.e., $\mathcal{T}(G) = \sum_{v \in V} \mathcal{T}(v)/\lvert V\vert $)~\cite{watts1998collective}.
Transitivity not only provides fundamental information on graphs, but also 
has been found to be a valuable metric in various applications, 
including neuroscience~\cite{masuda2018clustering, hsu2018applying, loeffler2020topological}, 
link prediction~\cite{chen2019application, wu2016link}, 
biology~\cite{kalna2007clustering, wang2011identification}, 
finance~\cite{tabak2014directed, cerqueti2021systemic}, 
web analysis~\cite{kutzkov2013streaming, becchetti2010efficient}, etc.

\begin{table}[t!]
    \vspace{-2mm}
        \begin{minipage}{0.22\textwidth}
		\centering
		\includegraphics[width=1.0\textwidth]{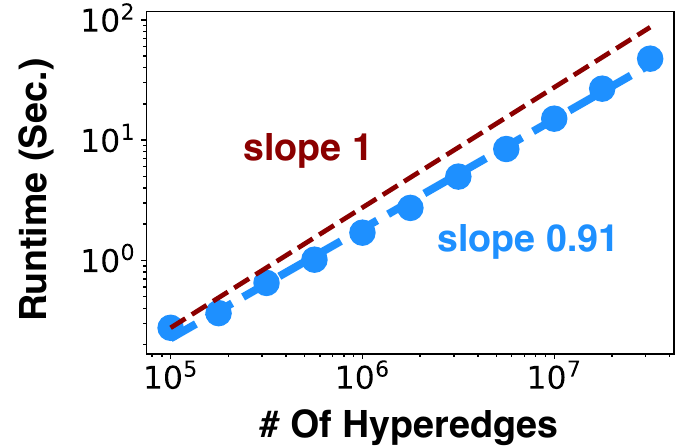}
		\captionof{figure}{Runtime of \generator, which is  linear in the number of hyperedges\label{fig:genlinear}.}
		\label{fig:runtime}
	\end{minipage}
        \hspace{3mm}
	\begin{minipage}{0.2\textwidth}
        \centering
		\label{tab:student}
        \scalebox{0.85}{
		\centering
		\begin{tabular}{cc}
			\toprule
			\textbf{Data}          & \textbf{Runtime (sec)} \\
			\midrule
			dblp     & 4.072      \\
			geology  & 2.393      \\
			history     & 1.858     \\
			\bottomrule
		\end{tabular}
        }
            \caption{Runtime of \generator when generating outputs for the real-world (\textit{coauth}) datasets with $\lvert V\vert > 10^{5}$. All take only a few seconds. \label{tab:genruntime}}
	\end{minipage}
    \vspace{-2mm}
\end{table}

\smallsection{Real-world hypergraph patterns} Hypergraphs are widely used in modeling group interactions of entities. 
Recently, there has been a focus on analyzing the patterns of real-world group interactions, {including
structural~\cite{do2020structural,ko2022growth,bu2023hypercore,juul2022hypergraph,tudisco2023core,lotito2022higher} and temporal \cite{benson2018sequences,lee2021thyme+,comrie2021hypergraph, cencetti2021temporal} properties, and especially, the
repetition \cite{benson2018sequences, choo2022persistence,cencetti2021temporal}, overlap~\cite{lee2021hyperedges,lee2020hypergraph,lee2021thyme+}, and reciprocity~\cite{kim2022reciprocity} of hyperedges.
Many of these patterns in real-world hypergraphs are not observed in random hypergraphs generated by null models.}

\smallsection{Transitivity in hypergraphs} Early attempts at measuring hypergraph transitivity were limited to the hypergraph level which only provided the overall extent of transitivity and could not give the local transitivity patterns in hypergraphs~\cite{estrada2005complex}. 
Recently, Behague et al.~\cite{behague2021iterated} addressed the unbounded (not lie in a fixed range) issue of a measure proposed by Estrada et al.~\cite{estrada2005complex}, but they also presented only global measures.
Local transitivity measures had been actively studied in the field of computational biology~\cite{gallagher2013clustering, zhou2011properties, klamt2009hypergraphs} since both hypergraphs and transitivity play important roles in modeling biological substances.
Additionally, Torres et al.~\cite{torres2021and} also suggested another way of quantifying local transitivity from the perspective of data mining.
However, these works mainly focused on binary relations and thus failed to distinguish the different degrees of intersection between groups.
Refer to Section~\ref{subsec:baselines} and the online appendix~\cite{github} for details of the limitations of the above measures.

\smallsection{Network generators reproducing real-world patterns}
Realistic network generation has been widely used for simulation, statistical testing, anonymization, and the upscaling of complex systems~\cite{leskovec2008random, nobari2011fast, lim2015survey, leskovec2010kronecker}.
Patterns in real-world networks, such as community structures~\cite{girvan2002community}, heavy-tailed node degree distribution~\cite{barabasi1999emergence, faloutsos1999power}, and {high} clustering coefficients~\cite{watts1998collective} have been reproduced by several network generators~\cite{leskovec2010kronecker, largeron2015generating, chakrabarti2004r} using simple and intuitive mechanisms.
Recently, there have been various attempts to reproduce the characteristics of real-world hypergraphs: 
{structural~\cite{do2020structural,giroire2022preferential} and temporal~\cite{ko2022growth} properties, and especially the repetition~\cite{benson2018sequences}, overlap~\cite{lee2021hyperedges}, and reciprocity~\cite{kim2022reciprocity} of hyperedges.
An efficient unified framework for hypergraph generators has also been proposed~\cite{hafner2022functional}.}

%% file: 007Conclusion.tex
In this work, we conduct a systematic and comprehensive analysis regarding the transitivity of real-world group {interactions}.
{We suggest seven properties of a well-defined hypergraph transitivity measure and propose \measure, which satisfies all these properties, with a fast computational algorithm \fastmeasure.}
By using \measure and \fastmeasure, we investigate the transitivity patterns of 12 real-world hypergraphs at four different levels.
Lastly, we propose \generator, a realistic and scalable hypergraph generator {that successfully reproduces these observed patterns.}

%% file: 100AppendixMotivation.tex
\subsection{Motivation and Necessity of Axioms}\label{subsec:axiommoti}

In this subsection, we provide the motivation and necessity that lie behind the axioms. They are designed to suggest \textbf{four desirable characteristics} of a measure with the following intuitions:

\smallsection{Boundness (\textsc{Axiom}~\ref{ax:hw_bounded} and ~\ref{ax:hg_bounded})} {Introducing a finite range for a measure provides an intuitive understanding} of the numerical extent of a characteristic. 
For example, if a measure does not lie in a fixed range, one cannot easily determine whether a certain hypergraph is transitive or not. 
Moreover, {a finite range enables meaningful comparisons between  different hypergraphs.} 
{Motivated by this fact}, we propose \textsc{Axiom}~\ref{ax:hw_bounded} and \textsc{Axiom}~\ref{ax:hg_bounded}, which suggest the bound of hyperwedge and hypergraph transitivity measures, respectively.

\smallsection{Extremal cases (\textsc{Axiom}~\ref{ax:min_hw_trans} and ~\ref{ax:max_hw_trans})} Gaining insight into when a measure achieves its maximum or minimum value {is crucial for understanding its behavior and interpreting its results effectively.}
{Thus}, we propose \textsc{Axiom}~\ref{ax:min_hw_trans} and \textsc{Axiom}~\ref{ax:max_hw_trans}, which describe cases where hyperwedge transitivity is minimized or maximized. 

\smallsection{Incremental changes (\textsc{Axiom}~\ref{ax:incre_changes_cands} and ~\ref{ax:incre_changes_replacement})} Understanding when the value of a measure increases (or decreases) is crucial for its interpretation and {to ensure its validity.}
Without this knowledge, one may distrust the measure, {and there is a risk of incorrect or incomplete interpretations of the measured value.}
Hence, we propose \textsc{Axiom}~\ref{ax:incre_changes_cands} and \textsc{Axiom}~\ref{ax:incre_changes_replacement}, which formalize when the measure increases.

\smallsection{Reducibility to pairwise-graph transitivity (\textsc{Axiom}~\ref{ax:reduce_pairwise_graph})} Transitivity in a graph is a well-known statistic that is widely used in various fields of study~\cite{watts1998collective, wasserman1994social}.
Since a hypergraph is a generalization of a graph, it is reasonable to expect that a hypergraph transitivity measure should be equivalent to the common graph transitivity measure when applied to any hypergraph that contains only size-2 hyperedges (i.e., graph, $\vert e\vert = 2, \forall e \in E$). 
Inspired by this motivation, we propose \textsc{Axiom}~\ref{ax:reduce_pairwise_graph}, which suggests this characteristic.
\vspace{-2mm}
\color{black}

\subsection{Reason of Axiom~\ref{ax:max_hw_trans}}\label{subsec:whynotheotherwayround}
{In this subsection, we clarify why we do not establish 
a necessary and sufficient (\textit{iff}) condition in
\textsc{Axiom}~\ref{ax:max_hw_trans}.}

The other way around of Axiom~\ref{ax:max_hw_trans} is equivalent to: if there exists a candidate hyperedge that includes both left and right wings, the transitivity of the corresponding hyperwedge should be maximized as 1 (\textsc{Axiom}~\ref{ax:hw_bounded}). 
Formally, 
$
\exists e \in C \ s.t.\ L(w) \cup R(w) \subseteq  e \Rightarrow \mathcal{T}(w,C) = 1
$.
{However, this is not universally applicable, as there are domains where it is justifiable for a measure to apply an additional penalty when a candidate hyperedge intersects with the body group of the hyperwedge~\cite{gallagher2013clustering}.} 
In such a case, if a candidate hyperedge includes all of $L(w)$, $R(w)$, and $B(w)$, its transitivity should not be equal to 1, despite it satisfies the condition of axiom. 
Thus, the axiom can not represent the required properties of the corresponding domain.

Then, can we replace $L(w) \cup R(w) \subseteq e$ in \textsc{Axiom}~\ref{ax:max_hw_trans} with $L(w) \cup R(w) = e$ and build an \textit{iff} condition (i.e., $\mathcal{T}(w,C) = 1 \Leftrightarrow \exists e \in C \ s.t. \ e \in L(w) \cup R(w)$)?
{However, in certain domains, it may not be appropriate for a measure to penalize external nodes (i.e., $V \setminus (L(w) \cup R(w))$).
In such cases, $\mathcal{T}(w,C) = 1 \Rightarrow L(w) \cup R(w) =e$ 
may not be well-suited, as the inclusion of external nodes prevents the measure from attaining maximum transitivity. 
{This limitation may restrict the applicability and usefulness of transitivity measures.}
} 



%% file: 101BaselineUsecase.tex
{In this section, we provide further explanations regarding three variants of \measure (\textbf{B7-9} in Section~\ref{subsec:theory}) and their limitations in usage. 
We will explore the potential for counterintuitive results that may arise from \textbf{B7-9} and the infeasibility they pose when comparing different hyperwedges.} Throughout our analyses, we focus on a hyperwedge $w$ with $L(w) = \{1,\cdots ,10\}$ and  $R(w) = \{11,\cdots ,20\}$, and a candidate hyperedge $e = \{1,\cdots, 19\}$.
{In addition, we assume the function $f$ defined in Eq~\eqref{eq:penalizeall}, and thus $f(w,e) = 0.9$.} 
\vspace{-2mm}
\subsection{Analyses of B7 and B8}

In analyses of \textbf{B7} and \textbf{B8}, we assume a scenario where $e$ is an existing candidate hyperedge, and $e'= \{10, 19, 20\}$ is newly added to the candidate set, where $f(w,e') = 0.02$. {More generally, we consider \textit{"new-coverage"} cases where an} additional candidate hyperedge includes nodes from both the left wing and right wing that were previously not included in the existing candidate hyperedges.\footnote{{Note that in \textit{new-coverage} cases, an increase in transitivity is expected. This is because the introduction of a new candidate hyperedge includes new interactions between the two wings that were previously not captured by any other candidate hyperedges. \label{footnote:shouldincrease}}}

\smallsection{B7 (Variant with {mean})} \textbf{B7} is a measure where the maximum operation over the scores of $\{v_{1},v_{2}\} \in P(w)$ {in \measure (i.e., Eq~\eqref{eq:hypertrans})} is replaced by the mean operation. 
According to \textbf{B7}, the transitivity of a hyperwedge may diminish in the case of \textit{new-coverage}, which is counterintuitive (see Footnote~\ref{footnote:shouldincrease}). 
{For example, adding $e'$ causes the transitivity of $w$ to decrease} from 0.81 to 0.41. 

\smallsection{B8. (Variant with {simple max})} \textbf{B8} {measures the maximum value of $f(w,e)$ among all $e \in C$.}
According to \textbf{B8},
the transitivity of a hyperwedge may remain the same in the case of \textit{new-coverage}, which is counterintuitive (see Footnote~\ref{footnote:shouldincrease}). 
For example, the transitivity of $w$ remains {the same at} 0.81, even with the addition of $e'$.

\input{main_complex}
\vspace{-2mm}
\subsection{Analysis of B9}
Below, we assume another hyperwedge $w'$ with $L(w') = \{1,2,3\}$ and $R(w') = \{4,5,6\}$. {The wings are much smaller than those of $w$.}

\smallsection{B9. (Variant without normalization)} {\textbf{B9} is a mesaure where the denominator in \measure (i.e., Eq~\eqref{eq:hypertrans}) is replaced by $1$,  resulting in the absence of normalization.}
Consequently, the value of a measure does not lie in a fixed range and heavily depends on the size of {hyperwedges}, making it infeasible to compare the transitivity of different hyperwedges.
{For example, the transitivity $\mathcal{T}(w, \{e\}; f)$ of $w$ is always greater than the transitivity $\mathcal{T}(w', \{e''\}; f)$ of $w'$ regardless of the choice of a candidate hyperedge $e''$.} 


\color{black}

%% file: main_complex.tex
\begin{table*}[t!]
\caption{Runtime (sec.) and memory consumption (MB) of five generators {for approximating} 12 real-world hypergraphs. 
{Note that \generator achieves the fastest generation time for all six hypergraphs with $\vert V \vert \geq 10^4$ (\textit{coauthorship} and \textit{q\&a}), and it exhibits the lowest memory consumption in four out of the six hypergraphs.
For each hypergraph, the best results are colored. `-' indicates that the generation process either exceeds the time limit of 12 hours or encounters an out-of-memory issue.}} \label{tab:gentimeandmemory}
    \centering
    \scalebox{0.75}{
    \renewcommand{\arraystretch}{0.92}
        \centering
        \begin{tabular}{c | c| c c c |c c c| c c |c c |c c | c}
            \toprule
            \multirow{2}{*}{\textbf{Statistic}} & \multirow{2}{*}{\textbf{Generator}} & \multicolumn{3}{c|}{coauthorship} & \multicolumn{3}{c|}{q\&a} & 
            \multicolumn{2}{c|}{email} & \multicolumn{2}{c|}{ndc} &  \multicolumn{2}{c|}{contact} & Average \\
            & & dblp & geology & history & ubuntu & math & server & enron & eu & classes & substances & high & primary & ranking \\
            \midrule
            \midrule
            \multirow{6}{*}{\makecell{\textbf{Runtime} \\ 
            \textbf{(sec.)}}} &\textbf{\generator} & \goat 4.07 & \goat 2.39 & \goat 1.86 & \goat 0.39 & \goat 0.33 & \goat 0.63 & 0.04 & 0.18 & 0.04 & \goat 0.09 & 0.10 & 0.18  & 1.7 \goat\\
            \cmidrule{2-15}
            
            &\textbf{HyperPA}~\cite{do2020structural} & - & - & - & 374.51 & - & - & 5.02 & 5011.84 & 1155.00 & - & 2.29 & 4.56 & 5.0\\
            
            &\textbf{HyperFF}~\cite{ko2022growth} & 226.79 & 114.24 & 53.48 & 12.78 & 17.23 & 3.15 & 0.05 & \goat 0.09 & 0.32 & 0.10 & \goat 0.02 & \goat 0.02 & 2.7\\
            
            &\textbf{HyperLap}~\cite{lee2021hyperedges} & 19.07 & 9.57 & 2.24 & 0.57 & 0.45 & 1.14 & \goat 0.01 & 0.14 & \goat 0.01 & 0.10 & 0.06 & 0.04 & 1.8\\

            &\textbf{HyperLap+}~\cite{lee2021hyperedges} & 1322.6 & 611.91 & 160.75 & 12.08 & 1.00 & 23.46 & 0.06 & 2.74 & 0.29 & 3.92 & 0.26 & 0.51 & 3.7\\
            \midrule 
            \midrule 
            
            \multirow{6}{*}{\makecell{\textbf{Memory} \\ \textbf{consumption} \\ 
            \textbf{(MB)}}} &\textbf{\generator} & \goattwo 1535 & \goattwo 761 & \goattwo 325 & 90 & \goattwo 36 & 129 & 2 & 3 & \goattwo 2 & 11 & 11 & 23 & 2.0 \goattwo\\
            \cmidrule{2-15}

            &\textbf{HyperPA}~\cite{do2020structural} & - & - & - & \goattwo 21 & - & - & 76 & 60169 & 21351 & - & 2 & 2 & 4.2\\
            
            &\textbf{HyperFF}~\cite{ko2022growth} & 3307 & 1655 & 739 & 107 & 147 & \goattwo 26 & \goattwo 1 & \goattwo 2 & 4 & \goattwo 1 & \goattwo 1 & \goattwo 1 & 2.0 \goattwo \\
            
            &\textbf{HyperLap}~\cite{lee2021hyperedges} & 3197 & 1498 & 529 & 134 & 86 & 204 & 11 & 32 & 15 & 41 & 12 & 11 & 3.1\\
            
            &\textbf{HyperLap+}~\cite{lee2021hyperedges} & 5042 & 3005 & 1901 & 591 & 110 & 321 & 9 & 39 & 14 & 412 & 10 & 18 & 3.8\\
            
            \bottomrule
        \end{tabular}
        }
\end{table*}

%% file: 102AppendixExperiment.tex
\begin{table}[t!]
	\caption{\label{tab:generatoranalysis} Complexity analysis of the generators.}
	\scalebox{0.76}{
		\begin{tabular}{c| c | c | c}
			\toprule
                \textbf{Generator} & \textbf{Time complexity} & \textbf{Memory complexity} & \textbf{Incremental} \\
			\midrule
                \textbf{\generator} & $\mathcal{O}(\log_{2}{\lvert V \vert }\times \sum_{e \in E} \lvert e \vert )$ & $\mathcal{O}(\lvert V\vert + \sum_{e \in E} \lvert e \vert)$ & \yes \\
                \midrule
			  \textbf{HyperPA}~\cite{do2020structural} & 
     $\mathcal{O}( \sum_{e \in E}\log_{2}{\binom{\lvert V\rvert}{\lvert e \vert}})$ & $\mathcal{O}(\sum_{e\in E}2^{\lvert e\vert})$ &  \yes \\
                \textbf{HyperFF}~\cite{ko2022growth} & $\mathcal{O}(\lvert V\vert \times \sum_{e \in E}  \lvert e \vert)  $ & $\mathcal{O}(\lvert V\vert + \sum_{e \in E} \lvert e \vert)$ &  \yes \\
                \textbf{HyperLap}~\cite{lee2021hyperedges} & $\mathcal{O}(\sum_{e \in E} \lvert e \vert )$ & $\mathcal{O}(\lvert V\vert + \sum_{e \in E} \lvert e \vert)$ &  \no \\
                \textbf{HyperLap+}~\cite{lee2021hyperedges} & $\mathcal{O}(\log_{2} {\lvert V\vert } \times \sum_{e \in E} \lvert e \vert )$ & $\mathcal{O}(\lvert V\vert + \sum_{e \in E} \lvert e \vert)$ &  \no \\
			\bottomrule
		\end{tabular}
  }
\end{table}

\subsection{Experimental Settings}\label{subsec:expsetting}

\smallsection{Machines and implementations} 
We use machines with Intel i9-10900K CPUs and 64GB RAM for all experiments.
We implement \naivemeasure, \fastmeasure, \naivegenerator, and \generator in Java 18.
{For all baseline generators, we use their implementations provided by the authors.}

\smallsection{Hyperparameters}
{We perform grid searches to fine-tune the hyperparameters aiming to minimize the difference between the (hypergraph-level) transitivity of the real-world hypergraphs and that of generated hypergraphs.
The search space of \generator is $\{0.5, 0.55, \cdots 0.9\}$ for $p$, $\{8, 9, \cdots, 15\}$ for $C$, and $\{2, 3, \cdots, 10\}$ for $\alpha$.
We set $\beta$ to
$2$ (if $\vert V\vert \leq 10^{4}$), $3$ (if $10^{4}< \vert V\vert \leq 10^{6}$), and $4$ (if $10^{6} < \vert V\vert$), depending on the number of nodes.
The search space of {HyperFF} is $\{0.49, 0.51\}$ for $p$ and $\{0.2, 0.3\}$ for $q$, as in \cite{ko2022growth}.
For {HyperLap}, as in \cite{lee2021hyperedges}, we use the uniform level distribution.
For {HyperLap+}, we tune $p$ within $\{0.01, 0.05, 0.1\}$, while it is fixed to $0.05$ in \cite{lee2021hyperedges}. 
{HyperPA} does not require any hyperparameters.}

\subsection{Additional Details and Results}\label{subsec:expresult}

We provide the details of the statistical test in Section~\ref{subsec:mainobservation} and conduct an empirical comparison of \generator with other hypergraph generators, with respect to additional hypergraph properties.

\smallsection{Statistical test} To demonstrate whether the differences between the transitivity of real-world hypergraphs and randomized hypergraphs are statistically significant, we conduct Z-tests using 10 randomized hypergraphs.
Specifically,
For each real-world hypergraph $G$, we create 10 randomized hypergraphs $\mathcal{G} = \{\mathcal{G}'_{1}, \cdots, \mathcal{G}'_{10}\}$ using {HyperCL}~\cite{lee2021hyperedges}.
{Then, we compute $Z = \frac{T(G) - \bar{T}(\mathcal{G})}{T_{sd}(\mathcal{G})/\sqrt{n}}$,
where $\bar{T}(\mathcal{G})$ and $T_{sd}(\mathcal{G})$ denote the average and standard deviation of the (hypergraph-level) transitivity values of the randomized hypergraphs.}
{See Table~\ref{tab:observation1} in the main paper for the results}.

\smallsection{Reproducibility of other properties} 
{In the main paper, our evaluation of hypergraph generators focuses on the transitivity patterns observed in real-world hypergraphs (Observation~\ref{obs:hypergraph}-\ref{obs:hyperedge}). 
Here, we compare the ability of \generator and the baseline generators to produce realistic hypergraphs with respect to various properties, using density, diameter, hyperedge size, degree, and intersection size.
We use the formulae in \cite{choe2022midas} for the statistics.}


{Table~\ref{tab:otherproperties} presents the average rank of each hypergraph generator across 12 real-world hypergraphs with respect to each property.
Among all five hypergraph generators, \generator achieves the highest rank with respect to density and the second-best rank with respect to diameter. Notably, among the incremental generators,\footnote{{Recall that, as discussed in Section~\ref{subsec:generatoreval}, incremental hypergraph generators create nodes and hyperedges incrementally, offering several advantages.}} \generator outperforms the others with respect to four out of five properties.} 

These results confirm that \generator exhibits competitive performance in replicating a wide range of hypergraph properties compared to other generators. For more detailed statistics, refer to the online appendix~\cite{github}.

\begin{table}[t]
	\caption{\label{tab:otherproperties} {Average rank of each hypergraph generator across 12 real-world datasets in terms of reproducing each property. 
 The ranks outside the parentheses indicate the ranks among all generators, and the ranks inside the parentheses indicate the ranks among the incremental generators only.
 The best results are in \textbf{bold} and the second best ones are \underline{underlined}.
 The prefix `H-' is used to indicate `Hyper-'.}}
	\scalebox{0.72}{
		\begin{tabular}{c| c | c c | c c }
			\toprule
                \multirow{2}{*}{\textbf{Generator}} & \multicolumn{3}{c|}{\textbf{Incremental generator}} & \multicolumn{2}{c}{\textbf{Static generator}}\\
                \cmidrule{2-6}
                 & \textbf{\generator} & \textbf{H-PA}~\cite{do2020structural} & \textbf{H-FF}~\cite{ko2022growth} & 
                 \textbf{H-Lap}~\cite{lee2021hyperedges} & \textbf{H-Lap+}~\cite{lee2021hyperedges} \\
			\midrule
	        Density & \textbf{1.25} (\textbf{1.00}) & 4.50 (2.67) & 4.08 (\underline{2.33}) & \underline{1.58} & 2.50 \\           		  
                Diameter & \underline{2.67} (\textbf{1.75}) & 4.08 (2.42) & 2.83 (\underline{1.83}) & \textbf{2.58} & 2.83 \\           		  
                Hyperedge size & 3.67 (\textbf{1.75}) & 3.92 (\underline{2.00}) & 4.08 (2.33) & \textbf{1.00} & \textbf{1.00} \\
                Degree & 3.91 (\underline{1.92}) & 3.93 (2.25) & 3.75 (\textbf{1.83}) & \textbf{1.25} & \underline{1.92} \\
                Intersection size & 3.00 (\textbf{1.67}) & 3.83 (2.42) & 3.08 (1.75) & \textbf{1.92} & \underline{2.42} \\
			\bottomrule
		\end{tabular}
  }
\end{table}

\subsection{Scalability Analysis}\label{subsec:appgenscale}

{We compare the scalability of \generator against other generators in terms of theoretical and empirical aspects.
Table~\ref{tab:generatoranalysis} demonstrates that \generator exhibits the lowest time and space complexity among the three incremental hypergraph generators. More detailed information can be found in the online appendix~\cite{github}.} 

{Furthermore, the empirical analysis of runtime and memory usage confirms the theoretical superiority of \generator.
We measure the runtime and memory consumption of each generator for approximating 12 real-world hypergraph datasets by generating hypergraphs of similar scale.
As shown in Table~\ref{tab:gentimeandmemory}, \generator demonstrates the shortest runtime among all generators for the six largest hypergraph datasets where $\vert V \vert > 10^{4}$ (coauthorship and q\&a). Additionally, \generator has the lowest memory consumption for four out of the six datasets.
Moreover, \generator attains the highest average rank in terms of both runtime and memory consumption across all datasets.
In summary, \generator is capable of generating large-scale hypergraphs with reduced runtime and memory requirements compared to other generators.}


%% file: main.bbl

\begin{thebibliography}{54}


\ifx \showCODEN    \undefined \def \showCODEN     #1{\unskip}     \fi
\ifx \showDOI      \undefined \def \showDOI       #1{#1}\fi
\ifx \showISBNx    \undefined \def \showISBNx     #1{\unskip}     \fi
\ifx \showISBNxiii \undefined \def \showISBNxiii  #1{\unskip}     \fi
\ifx \showISSN     \undefined \def \showISSN      #1{\unskip}     \fi
\ifx \showLCCN     \undefined \def \showLCCN      #1{\unskip}     \fi
\ifx \shownote     \undefined \def \shownote      #1{#1}          \fi
\ifx \showarticletitle \undefined \def \showarticletitle #1{#1}   \fi
\ifx \showURL      \undefined \def \showURL       {\relax}        \fi
\providecommand\bibfield[2]{#2}
\providecommand\bibinfo[2]{#2}
\providecommand\natexlab[1]{#1}
\providecommand\showeprint[2][]{arXiv:#2}

\bibitem[Barab{\'a}si and Albert(1999)]%
        {barabasi1999emergence}
\bibfield{author}{\bibinfo{person}{Albert-L{\'a}szl{\'o} Barab{\'a}si} {and}
  \bibinfo{person}{R{\'e}ka Albert}.} \bibinfo{year}{1999}\natexlab{}.
\newblock \showarticletitle{Emergence of scaling in random networks}.
\newblock \bibinfo{journal}{\emph{Science}} \bibinfo{volume}{286},
  \bibinfo{number}{5439} (\bibinfo{year}{1999}), \bibinfo{pages}{509--512}.
\newblock


\bibitem[Barrat et~al\mbox{.}(2004)]%
        {barrat2004architecture}
\bibfield{author}{\bibinfo{person}{Alain Barrat}, \bibinfo{person}{Marc
  Barthelemy}, \bibinfo{person}{Romualdo Pastor-Satorras}, {and}
  \bibinfo{person}{Alessandro Vespignani}.} \bibinfo{year}{2004}\natexlab{}.
\newblock \showarticletitle{The architecture of complex weighted networks}.
\newblock \bibinfo{journal}{\emph{Proceedings of the National Academy of
  Sciences}} \bibinfo{volume}{101}, \bibinfo{number}{11}
  (\bibinfo{year}{2004}), \bibinfo{pages}{3747--3752}.
\newblock


\bibitem[Becchetti et~al\mbox{.}(2010)]%
        {becchetti2010efficient}
\bibfield{author}{\bibinfo{person}{Luca Becchetti}, \bibinfo{person}{Paolo
  Boldi}, \bibinfo{person}{Carlos Castillo}, {and} \bibinfo{person}{Aristides
  Gionis}.} \bibinfo{year}{2010}\natexlab{}.
\newblock \showarticletitle{Efficient algorithms for large-scale local triangle
  counting}.
\newblock \bibinfo{journal}{\emph{ACM Transactions on Knowledge Discovery from
  Data}} \bibinfo{volume}{4}, \bibinfo{number}{3} (\bibinfo{year}{2010}),
  \bibinfo{pages}{1--28}.
\newblock


\bibitem[Behague et~al\mbox{.}(2023)]%
        {behague2021iterated}
\bibfield{author}{\bibinfo{person}{Natalie~C Behague}, \bibinfo{person}{Anthony
  Bonato}, \bibinfo{person}{Melissa~A Huggan}, \bibinfo{person}{Rehan Malik},
  {and} \bibinfo{person}{Trent~G Marbach}.} \bibinfo{year}{2023}\natexlab{}.
\newblock \showarticletitle{The iterated local transitivity model for
  hypergraphs}.
\newblock \bibinfo{journal}{\emph{Discrete Applied Mathematics}}
  \bibinfo{volume}{337} (\bibinfo{year}{2023}), \bibinfo{pages}{106--119}.
\newblock


\bibitem[Benson et~al\mbox{.}(2018a)]%
        {benson2018simplicial}
\bibfield{author}{\bibinfo{person}{Austin~R Benson}, \bibinfo{person}{Rediet
  Abebe}, \bibinfo{person}{Michael~T Schaub}, \bibinfo{person}{Ali Jadbabaie},
  {and} \bibinfo{person}{Jon Kleinberg}.} \bibinfo{year}{2018}\natexlab{a}.
\newblock \showarticletitle{Simplicial closure and higher-order link
  prediction}.
\newblock \bibinfo{journal}{\emph{Proceedings of the National Academy of
  Sciences}} \bibinfo{volume}{115}, \bibinfo{number}{48}
  (\bibinfo{year}{2018}), \bibinfo{pages}{E11221--E11230}.
\newblock


\bibitem[Benson et~al\mbox{.}(2018b)]%
        {benson2018sequences}
\bibfield{author}{\bibinfo{person}{Austin~R Benson}, \bibinfo{person}{Ravi
  Kumar}, {and} \bibinfo{person}{Andrew Tomkins}.}
  \bibinfo{year}{2018}\natexlab{b}.
\newblock \showarticletitle{Sequences of sets}. In
  \bibinfo{booktitle}{\emph{KDD}}.
\newblock


\bibitem[Bringmann and Panagiotou(2012)]%
        {bringmann2012efficient}
\bibfield{author}{\bibinfo{person}{Karl Bringmann} {and}
  \bibinfo{person}{Konstantinos Panagiotou}.} \bibinfo{year}{2012}\natexlab{}.
\newblock \showarticletitle{Efficient sampling methods for discrete
  distributions}. In \bibinfo{booktitle}{\emph{ICALP}}.
\newblock


\bibitem[Bu et~al\mbox{.}(2023)]%
        {bu2023hypercore}
\bibfield{author}{\bibinfo{person}{Fanchen Bu}, \bibinfo{person}{Geon Lee},
  {and} \bibinfo{person}{Kijung Shin}.} \bibinfo{year}{2023}\natexlab{}.
\newblock \showarticletitle{Hypercore Decomposition for Non-Fragile Hyperedges:
  Concepts, Algorithms, Observations, and Applications}.
\newblock \bibinfo{journal}{\emph{ArXiv}} (\bibinfo{year}{2023}).
\newblock


\bibitem[Cencetti et~al\mbox{.}(2021)]%
        {cencetti2021temporal}
\bibfield{author}{\bibinfo{person}{Giulia Cencetti}, \bibinfo{person}{Federico
  Battiston}, \bibinfo{person}{Bruno Lepri}, {and} \bibinfo{person}{M{\'a}rton
  Karsai}.} \bibinfo{year}{2021}\natexlab{}.
\newblock \showarticletitle{Temporal properties of higher-order interactions in
  social networks}.
\newblock \bibinfo{journal}{\emph{Scientific reports}} \bibinfo{volume}{11},
  \bibinfo{number}{1} (\bibinfo{year}{2021}), \bibinfo{pages}{1--10}.
\newblock


\bibitem[Cerqueti et~al\mbox{.}(2021)]%
        {cerqueti2021systemic}
\bibfield{author}{\bibinfo{person}{Roy Cerqueti}, \bibinfo{person}{Gian~Paolo
  Clemente}, {and} \bibinfo{person}{Rosanna Grassi}.}
  \bibinfo{year}{2021}\natexlab{}.
\newblock \showarticletitle{Systemic risk assessment through high order
  clustering coefficient}.
\newblock \bibinfo{journal}{\emph{Annals of Operations Research}}
  \bibinfo{volume}{299}, \bibinfo{number}{1} (\bibinfo{year}{2021}),
  \bibinfo{pages}{1165--1187}.
\newblock


\bibitem[Chakrabarti et~al\mbox{.}(2004)]%
        {chakrabarti2004r}
\bibfield{author}{\bibinfo{person}{Deepayan Chakrabarti},
  \bibinfo{person}{Yiping Zhan}, {and} \bibinfo{person}{Christos Faloutsos}.}
  \bibinfo{year}{2004}\natexlab{}.
\newblock \showarticletitle{R-MAT: A recursive model for graph mining}. In
  \bibinfo{booktitle}{\emph{SDM}}.
\newblock


\bibitem[Chen et~al\mbox{.}(2019)]%
        {chen2019application}
\bibfield{author}{\bibinfo{person}{Xing Chen}, \bibinfo{person}{Ling Fang},
  \bibinfo{person}{Tinghong Yang}, \bibinfo{person}{Jian Yang},
  \bibinfo{person}{Zerong Bao}, \bibinfo{person}{Duzhi Wu}, {and}
  \bibinfo{person}{Jing Zhao}.} \bibinfo{year}{2019}\natexlab{}.
\newblock \showarticletitle{The application of degree related clustering
  coefficient in estimating the link predictability and predicting missing
  links of networks}.
\newblock \bibinfo{journal}{\emph{Chaos: An Interdisciplinary Journal of
  Nonlinear Science}} \bibinfo{volume}{29}, \bibinfo{number}{5}
  (\bibinfo{year}{2019}), \bibinfo{pages}{053135}.
\newblock


\bibitem[Choe et~al\mbox{.}(2022)]%
        {choe2022midas}
\bibfield{author}{\bibinfo{person}{Minyoung Choe}, \bibinfo{person}{Jaemin
  Yoo}, \bibinfo{person}{Geon Lee}, \bibinfo{person}{Woonsung Baek},
  \bibinfo{person}{U Kang}, {and} \bibinfo{person}{Kijung Shin}.}
  \bibinfo{year}{2022}\natexlab{}.
\newblock \showarticletitle{Midas: Representative sampling from real-world
  hypergraphs}. In \bibinfo{booktitle}{\emph{WWW}}.
\newblock


\bibitem[Choo and Shin(2022)]%
        {choo2022persistence}
\bibfield{author}{\bibinfo{person}{Hyunjin Choo} {and} \bibinfo{person}{Kijung
  Shin}.} \bibinfo{year}{2022}\natexlab{}.
\newblock \showarticletitle{On the persistence of higher-order interactions in
  real-world hypergraphs}. In \bibinfo{booktitle}{\emph{SDM}}.
\newblock


\bibitem[Comrie and Kleinberg(2021)]%
        {comrie2021hypergraph}
\bibfield{author}{\bibinfo{person}{Cazamere Comrie} {and} \bibinfo{person}{Jon
  Kleinberg}.} \bibinfo{year}{2021}\natexlab{}.
\newblock \showarticletitle{Hypergraph Ego-networks and Their Temporal
  Evolution}. In \bibinfo{booktitle}{\emph{ICDM}}.
\newblock


\bibitem[Do et~al\mbox{.}(2020)]%
        {do2020structural}
\bibfield{author}{\bibinfo{person}{Manh~Tuan Do}, \bibinfo{person}{Se-eun
  Yoon}, \bibinfo{person}{Bryan Hooi}, {and} \bibinfo{person}{Kijung Shin}.}
  \bibinfo{year}{2020}\natexlab{}.
\newblock \showarticletitle{Structural patterns and generative models of
  real-world hypergraphs}. In \bibinfo{booktitle}{\emph{KDD}}.
\newblock


\bibitem[Estrada and Rodriguez-Velazquez(2005)]%
        {estrada2005complex}
\bibfield{author}{\bibinfo{person}{Ernesto Estrada} {and}
  \bibinfo{person}{Juan~A Rodriguez-Velazquez}.}
  \bibinfo{year}{2005}\natexlab{}.
\newblock \showarticletitle{Complex networks as hypergraphs}.
\newblock \bibinfo{journal}{\emph{ArXiv}} (\bibinfo{year}{2005}).
\newblock


\bibitem[Faloutsos et~al\mbox{.}(1999)]%
        {faloutsos1999power}
\bibfield{author}{\bibinfo{person}{Michalis Faloutsos}, \bibinfo{person}{Petros
  Faloutsos}, {and} \bibinfo{person}{Christos Faloutsos}.}
  \bibinfo{year}{1999}\natexlab{}.
\newblock \showarticletitle{On power-law relationships of the internet
  topology}.
\newblock \bibinfo{journal}{\emph{ACM SIGCOMM computer communication review}}
  \bibinfo{volume}{29}, \bibinfo{number}{4} (\bibinfo{year}{1999}),
  \bibinfo{pages}{251--262}.
\newblock


\bibitem[Fieller et~al\mbox{.}(1957)]%
        {fieller1957tests}
\bibfield{author}{\bibinfo{person}{Edgar~C Fieller}, \bibinfo{person}{Herman~O
  Hartley}, {and} \bibinfo{person}{Egon~S Pearson}.}
  \bibinfo{year}{1957}\natexlab{}.
\newblock \showarticletitle{Tests for rank correlation coefficients. I}.
\newblock \bibinfo{journal}{\emph{Biometrika}} \bibinfo{volume}{44},
  \bibinfo{number}{3/4} (\bibinfo{year}{1957}), \bibinfo{pages}{470--481}.
\newblock


\bibitem[Gallagher and Goldberg(2013)]%
        {gallagher2013clustering}
\bibfield{author}{\bibinfo{person}{Suzanne~Renick Gallagher} {and}
  \bibinfo{person}{Debra~S Goldberg}.} \bibinfo{year}{2013}\natexlab{}.
\newblock \showarticletitle{Clustering coefficients in protein interaction
  hypernetworks}. In \bibinfo{booktitle}{\emph{BCB}}.
\newblock


\bibitem[Giroire et~al\mbox{.}(2022)]%
        {giroire2022preferential}
\bibfield{author}{\bibinfo{person}{Fr{\'e}d{\'e}ric Giroire},
  \bibinfo{person}{Nicolas Nisse}, \bibinfo{person}{Thibaud Trolliet}, {and}
  \bibinfo{person}{Ma{\l}gorzata Sulkowska}.} \bibinfo{year}{2022}\natexlab{}.
\newblock \showarticletitle{Preferential attachment hypergraph with high
  modularity}.
\newblock \bibinfo{journal}{\emph{Network Science}} \bibinfo{volume}{10},
  \bibinfo{number}{4} (\bibinfo{year}{2022}), \bibinfo{pages}{400--429}.
\newblock


\bibitem[Girvan and Newman(2002)]%
        {girvan2002community}
\bibfield{author}{\bibinfo{person}{Michelle Girvan} {and}
  \bibinfo{person}{Mark~EJ Newman}.} \bibinfo{year}{2002}\natexlab{}.
\newblock \showarticletitle{Community structure in social and biological
  networks}.
\newblock \bibinfo{journal}{\emph{Proceedings of the National Academy of
  Sciences}} \bibinfo{volume}{99}, \bibinfo{number}{12} (\bibinfo{year}{2002}),
  \bibinfo{pages}{7821--7826}.
\newblock


\bibitem[Hafner et~al\mbox{.}(2022)]%
        {hafner2022functional}
\bibfield{author}{\bibinfo{person}{Lilith~Orion Hafner}, \bibinfo{person}{Chase
  Holdener}, {and} \bibinfo{person}{Nicole Eikmeier}.}
  \bibinfo{year}{2022}\natexlab{}.
\newblock \showarticletitle{Functional Ball Dropping: A superfast hypergraph
  generation scheme}. In \bibinfo{booktitle}{\emph{BigData}}.
\newblock


\bibitem[Hsu et~al\mbox{.}(2018)]%
        {hsu2018applying}
\bibfield{author}{\bibinfo{person}{Chen-Fang Hsu}, \bibinfo{person}{Tsair-Wei
  Chien}, \bibinfo{person}{Julie~Chi Chow}, {and} \bibinfo{person}{Willy
  Chou}.} \bibinfo{year}{2018}\natexlab{}.
\newblock \showarticletitle{Applying clustering coefficient to the pattern of
  international author collaboration in neuroimmunology and neuroinflammation}.
\newblock \bibinfo{journal}{\emph{Neuroimmunology and Neuroinflammation}}
  \bibinfo{volume}{5} (\bibinfo{year}{2018}), \bibinfo{pages}{9}.
\newblock


\bibitem[Juul et~al\mbox{.}(2022)]%
        {juul2022hypergraph}
\bibfield{author}{\bibinfo{person}{Jonas~L Juul}, \bibinfo{person}{Austin~R
  Benson}, {and} \bibinfo{person}{Jon Kleinberg}.}
  \bibinfo{year}{2022}\natexlab{}.
\newblock \showarticletitle{Hypergraph patterns and collaboration structure}.
\newblock \bibinfo{journal}{\emph{arXiv preprint arXiv:2210.02163}}
  (\bibinfo{year}{2022}).
\newblock


\bibitem[Kalna and Higham(2007)]%
        {kalna2007clustering}
\bibfield{author}{\bibinfo{person}{Gabriela Kalna} {and}
  \bibinfo{person}{Desmond~J Higham}.} \bibinfo{year}{2007}\natexlab{}.
\newblock \showarticletitle{A clustering coefficient for weighted networks,
  with application to gene expression data}.
\newblock \bibinfo{journal}{\emph{AI Communications}} \bibinfo{volume}{20},
  \bibinfo{number}{4} (\bibinfo{year}{2007}), \bibinfo{pages}{263--271}.
\newblock


\bibitem[Kim et~al\mbox{.}(2023)]%
        {github}
\bibfield{author}{\bibinfo{person}{Sunwoo Kim}, \bibinfo{person}{Fanchen Bu},
  \bibinfo{person}{Minyoung Choe}, \bibinfo{person}{Jaemin Yoo}, {and}
  \bibinfo{person}{Kijung Shin}.} \bibinfo{year}{2023}\natexlab{}.
\newblock \showarticletitle{How Transitive Are Real-World Group Interactions? -
  Measurement and Reproduction (Code, Datasets, and Online Appendix)}.
\newblock
\urldef\tempurl%
\url{https://github.com/kswoo97/hypertrans}
\showURL{%
\tempurl}


\bibitem[Kim et~al\mbox{.}(2022)]%
        {kim2022reciprocity}
\bibfield{author}{\bibinfo{person}{Sunwoo Kim}, \bibinfo{person}{Minyoung
  Choe}, \bibinfo{person}{Jaemin Yoo}, {and} \bibinfo{person}{Kijung Shin}.}
  \bibinfo{year}{2022}\natexlab{}.
\newblock \showarticletitle{Reciprocity in Directed Hypergraphs: Measures,
  Findings, and Generators}. In \bibinfo{booktitle}{\emph{ICDM}}.
\newblock


\bibitem[Klamt et~al\mbox{.}(2009)]%
        {klamt2009hypergraphs}
\bibfield{author}{\bibinfo{person}{Steffen Klamt}, \bibinfo{person}{Utz-Uwe
  Haus}, {and} \bibinfo{person}{Fabian Theis}.}
  \bibinfo{year}{2009}\natexlab{}.
\newblock \showarticletitle{Hypergraphs and cellular networks}.
\newblock \bibinfo{journal}{\emph{Plos Computational Biology}}
  \bibinfo{volume}{5}, \bibinfo{number}{5} (\bibinfo{year}{2009}),
  \bibinfo{pages}{e1000385}.
\newblock


\bibitem[Ko et~al\mbox{.}(2022)]%
        {ko2022growth}
\bibfield{author}{\bibinfo{person}{Jihoon Ko}, \bibinfo{person}{Yunbum Kook},
  {and} \bibinfo{person}{Kijung Shin}.} \bibinfo{year}{2022}\natexlab{}.
\newblock \showarticletitle{Growth patterns and models of real-world
  hypergraphs}.
\newblock \bibinfo{journal}{\emph{Knowledge and Information Systems}}
  \bibinfo{volume}{64}, \bibinfo{number}{11} (\bibinfo{year}{2022}),
  \bibinfo{pages}{2883--2920}.
\newblock


\bibitem[Kutzkov and Pagh(2013)]%
        {kutzkov2013streaming}
\bibfield{author}{\bibinfo{person}{Konstantin Kutzkov} {and}
  \bibinfo{person}{Rasmus Pagh}.} \bibinfo{year}{2013}\natexlab{}.
\newblock \showarticletitle{On the streaming complexity of computing local
  clustering coefficients}. In \bibinfo{booktitle}{\emph{WSDM}}.
\newblock


\bibitem[Largeron et~al\mbox{.}(2015)]%
        {largeron2015generating}
\bibfield{author}{\bibinfo{person}{Christine Largeron},
  \bibinfo{person}{Pierre-Nicolas Mougel}, \bibinfo{person}{Reihaneh Rabbany},
  {and} \bibinfo{person}{Osmar~R Za{\"\i}ane}.}
  \bibinfo{year}{2015}\natexlab{}.
\newblock \showarticletitle{Generating attributed networks with communities}.
\newblock \bibinfo{journal}{\emph{Plos One}} \bibinfo{volume}{10},
  \bibinfo{number}{4} (\bibinfo{year}{2015}), \bibinfo{pages}{e0122777}.
\newblock


\bibitem[Lee et~al\mbox{.}(2021)]%
        {lee2021hyperedges}
\bibfield{author}{\bibinfo{person}{Geon Lee}, \bibinfo{person}{Minyoung Choe},
  {and} \bibinfo{person}{Kijung Shin}.} \bibinfo{year}{2021}\natexlab{}.
\newblock \showarticletitle{How do hyperedges overlap in real-world
  hypergraphs?-patterns, measures, and generators}. In
  \bibinfo{booktitle}{\emph{WWW}}.
\newblock


\bibitem[Lee et~al\mbox{.}(2020)]%
        {lee2020hypergraph}
\bibfield{author}{\bibinfo{person}{Geon Lee}, \bibinfo{person}{Jihoon Ko},
  {and} \bibinfo{person}{Kijung Shin}.} \bibinfo{year}{2020}\natexlab{}.
\newblock \showarticletitle{Hypergraph motifs: concepts, algorithms, and
  discoveries}.
\newblock \bibinfo{journal}{\emph{PVLDB}} \bibinfo{volume}{13},
  \bibinfo{number}{12} (\bibinfo{year}{2020}), \bibinfo{pages}{2256--2269}.
\newblock


\bibitem[Lee and Shin(2021)]%
        {lee2021thyme+}
\bibfield{author}{\bibinfo{person}{Geon Lee} {and} \bibinfo{person}{Kijung
  Shin}.} \bibinfo{year}{2021}\natexlab{}.
\newblock \showarticletitle{Thyme+: Temporal hypergraph motifs and fast
  algorithms for exact counting}. In \bibinfo{booktitle}{\emph{ICDM}}.
\newblock


\bibitem[Leskovec(2008)]%
        {leskovec2008random}
\bibfield{author}{\bibinfo{person}{Jurij Leskovec}.}
  \bibinfo{year}{2008}\natexlab{}.
\newblock \showarticletitle{Dynamics of large networks}. In
  \bibinfo{booktitle}{\emph{Carnegie Mellon University}}.
\newblock


\bibitem[Leskovec et~al\mbox{.}(2010)]%
        {leskovec2010kronecker}
\bibfield{author}{\bibinfo{person}{Jure Leskovec}, \bibinfo{person}{Deepayan
  Chakrabarti}, \bibinfo{person}{Jon Kleinberg}, \bibinfo{person}{Christos
  Faloutsos}, {and} \bibinfo{person}{Zoubin Ghahramani}.}
  \bibinfo{year}{2010}\natexlab{}.
\newblock \showarticletitle{Kronecker graphs: an approach to modeling
  networks.}
\newblock \bibinfo{journal}{\emph{Journal of Machine Learning Research}}
  \bibinfo{volume}{11}, \bibinfo{number}{2} (\bibinfo{year}{2010}).
\newblock


\bibitem[Lim et~al\mbox{.}(2015)]%
        {lim2015survey}
\bibfield{author}{\bibinfo{person}{Seung-Hwan Lim},
  \bibinfo{person}{Sangkeun~Matt Lee}, \bibinfo{person}{Sarah Powers},
  \bibinfo{person}{Mallikarjun Shankar}, {and} \bibinfo{person}{Neena Imam}.}
  \bibinfo{year}{2015}\natexlab{}.
\newblock \showarticletitle{Survey of approaches to generate realistic
  synthetic graphs}.
\newblock \bibinfo{journal}{\emph{Oak Ridge National Laboratory}}
  (\bibinfo{year}{2015}).
\newblock


\bibitem[Loeffler et~al\mbox{.}(2020)]%
        {loeffler2020topological}
\bibfield{author}{\bibinfo{person}{Alon Loeffler}, \bibinfo{person}{Ruomin
  Zhu}, \bibinfo{person}{Joel Hochstetter}, \bibinfo{person}{Mike Li},
  \bibinfo{person}{Kaiwei Fu}, \bibinfo{person}{Adrian Diaz-Alvarez},
  \bibinfo{person}{Tomonobu Nakayama}, \bibinfo{person}{James~M Shine}, {and}
  \bibinfo{person}{Zdenka Kuncic}.} \bibinfo{year}{2020}\natexlab{}.
\newblock \showarticletitle{Topological properties of neuromorphic nanowire
  networks}.
\newblock \bibinfo{journal}{\emph{Frontiers in Neuroscience}}
  \bibinfo{volume}{14} (\bibinfo{year}{2020}), \bibinfo{pages}{184}.
\newblock


\bibitem[Lotito et~al\mbox{.}(2022)]%
        {lotito2022higher}
\bibfield{author}{\bibinfo{person}{Quintino~Francesco Lotito},
  \bibinfo{person}{Federico Musciotto}, \bibinfo{person}{Alberto Montresor},
  {and} \bibinfo{person}{Federico Battiston}.} \bibinfo{year}{2022}\natexlab{}.
\newblock \showarticletitle{Higher-order motif analysis in hypergraphs}.
\newblock \bibinfo{journal}{\emph{Communications Physics}} \bibinfo{volume}{5},
  \bibinfo{number}{1} (\bibinfo{year}{2022}), \bibinfo{pages}{1--8}.
\newblock


\bibitem[Masuda et~al\mbox{.}(2018)]%
        {masuda2018clustering}
\bibfield{author}{\bibinfo{person}{Naoki Masuda}, \bibinfo{person}{Michiko
  Sakaki}, \bibinfo{person}{Takahiro Ezaki}, {and} \bibinfo{person}{Takamitsu
  Watanabe}.} \bibinfo{year}{2018}\natexlab{}.
\newblock \showarticletitle{Clustering coefficients for correlation networks}.
\newblock \bibinfo{journal}{\emph{Frontiers in Neuroinformatics}}
  \bibinfo{volume}{12} (\bibinfo{year}{2018}), \bibinfo{pages}{7}.
\newblock


\bibitem[Newman et~al\mbox{.}(2001)]%
        {newman2001random}
\bibfield{author}{\bibinfo{person}{Mark~EJ Newman}, \bibinfo{person}{Steven~H
  Strogatz}, {and} \bibinfo{person}{Duncan~J Watts}.}
  \bibinfo{year}{2001}\natexlab{}.
\newblock \showarticletitle{Random graphs with arbitrary degree distributions
  and their applications}.
\newblock \bibinfo{journal}{\emph{Physical review E}} \bibinfo{volume}{64},
  \bibinfo{number}{2} (\bibinfo{year}{2001}), \bibinfo{pages}{026118}.
\newblock


\bibitem[Nobari et~al\mbox{.}(2011)]%
        {nobari2011fast}
\bibfield{author}{\bibinfo{person}{Sadegh Nobari}, \bibinfo{person}{Xuesong
  Lu}, \bibinfo{person}{Panagiotis Karras}, {and} \bibinfo{person}{St{\'e}phane
  Bressan}.} \bibinfo{year}{2011}\natexlab{}.
\newblock \showarticletitle{Fast random graph generation}. In
  \bibinfo{booktitle}{\emph{EDBT}}.
\newblock


\bibitem[Ravasz and Barab{\'a}si(2003)]%
        {ravasz2003hierarchical}
\bibfield{author}{\bibinfo{person}{Erzs{\'e}bet Ravasz} {and}
  \bibinfo{person}{Albert-L{\'a}szl{\'o} Barab{\'a}si}.}
  \bibinfo{year}{2003}\natexlab{}.
\newblock \showarticletitle{Hierarchical organization in complex networks}.
\newblock \bibinfo{journal}{\emph{Physical Review E}} \bibinfo{volume}{67},
  \bibinfo{number}{2} (\bibinfo{year}{2003}), \bibinfo{pages}{026112}.
\newblock


\bibitem[Schwarz(2011)]%
        {schwarz2011darts}
\bibfield{author}{\bibinfo{person}{Keith Schwarz}.}
  \bibinfo{year}{2011}\natexlab{}.
\newblock \showarticletitle{Darts, dice, and coins: Sampling from a discrete
  distribution}.
\newblock \bibinfo{journal}{\emph{Retrieved}} \bibinfo{volume}{3},
  \bibinfo{number}{28} (\bibinfo{year}{2011}), \bibinfo{pages}{2012}.
\newblock


\bibitem[Tabak et~al\mbox{.}(2014)]%
        {tabak2014directed}
\bibfield{author}{\bibinfo{person}{Benjamin~M Tabak}, \bibinfo{person}{Marcelo
  Takami}, \bibinfo{person}{Jadson~MC Rocha}, \bibinfo{person}{Daniel~O
  Cajueiro}, {and} \bibinfo{person}{Sergio~RS Souza}.}
  \bibinfo{year}{2014}\natexlab{}.
\newblock \showarticletitle{Directed clustering coefficient as a measure of
  systemic risk in complex banking networks}.
\newblock \bibinfo{journal}{\emph{Physica A: Statistical Mechanics and its
  Applications}}  \bibinfo{volume}{394} (\bibinfo{year}{2014}),
  \bibinfo{pages}{211--216}.
\newblock


\bibitem[Torres et~al\mbox{.}(2021)]%
        {torres2021and}
\bibfield{author}{\bibinfo{person}{Leo Torres}, \bibinfo{person}{Ann~S
  Blevins}, \bibinfo{person}{Danielle Bassett}, {and} \bibinfo{person}{Tina
  Eliassi-Rad}.} \bibinfo{year}{2021}\natexlab{}.
\newblock \showarticletitle{The why, how, and when of representations for
  complex systems}.
\newblock \bibinfo{journal}{\emph{SIAM Rev.}} \bibinfo{volume}{63},
  \bibinfo{number}{3} (\bibinfo{year}{2021}), \bibinfo{pages}{435--485}.
\newblock


\bibitem[Tudisco and Higham(2023)]%
        {tudisco2023core}
\bibfield{author}{\bibinfo{person}{Francesco Tudisco} {and}
  \bibinfo{person}{Desmond~J Higham}.} \bibinfo{year}{2023}\natexlab{}.
\newblock \showarticletitle{Core-periphery detection in hypergraphs}.
\newblock \bibinfo{journal}{\emph{SIAM Journal on Mathematics of Data Science}}
  \bibinfo{volume}{5}, \bibinfo{number}{1} (\bibinfo{year}{2023}),
  \bibinfo{pages}{1--21}.
\newblock


\bibitem[Wang et~al\mbox{.}(2011)]%
        {wang2011identification}
\bibfield{author}{\bibinfo{person}{Jianxin Wang}, \bibinfo{person}{Min Li},
  \bibinfo{person}{Huan Wang}, {and} \bibinfo{person}{Yi Pan}.}
  \bibinfo{year}{2011}\natexlab{}.
\newblock \showarticletitle{Identification of essential proteins based on edge
  clustering coefficient}.
\newblock \bibinfo{journal}{\emph{IEEE/ACM Transactions on Computational
  Biology and Bioinformatics}} \bibinfo{volume}{9}, \bibinfo{number}{4}
  (\bibinfo{year}{2011}), \bibinfo{pages}{1070--1080}.
\newblock


\bibitem[Wasserman et~al\mbox{.}(1994)]%
        {wasserman1994social}
\bibfield{author}{\bibinfo{person}{Stanley Wasserman},
  \bibinfo{person}{Katherine Faust}, {et~al\mbox{.}}}
  \bibinfo{year}{1994}\natexlab{}.
\newblock \showarticletitle{Social network analysis: Methods and applications}.
\newblock  (\bibinfo{year}{1994}).
\newblock


\bibitem[Watts and Strogatz(1998)]%
        {watts1998collective}
\bibfield{author}{\bibinfo{person}{Duncan~J Watts} {and}
  \bibinfo{person}{Steven~H Strogatz}.} \bibinfo{year}{1998}\natexlab{}.
\newblock \showarticletitle{Collective dynamics of ‘small-world’networks}.
\newblock \bibinfo{journal}{\emph{Nature}} \bibinfo{volume}{393},
  \bibinfo{number}{6684} (\bibinfo{year}{1998}), \bibinfo{pages}{440--442}.
\newblock


\bibitem[Wu et~al\mbox{.}(2016)]%
        {wu2016link}
\bibfield{author}{\bibinfo{person}{Zhihao Wu}, \bibinfo{person}{Youfang Lin},
  \bibinfo{person}{Jing Wang}, {and} \bibinfo{person}{Steve Gregory}.}
  \bibinfo{year}{2016}\natexlab{}.
\newblock \showarticletitle{Link prediction with node clustering coefficient}.
\newblock \bibinfo{journal}{\emph{Physica A: Statistical Mechanics and its
  Applications}}  \bibinfo{volume}{452} (\bibinfo{year}{2016}),
  \bibinfo{pages}{1--8}.
\newblock


\bibitem[Zhou et~al\mbox{.}(2005)]%
        {zhou2005maximal}
\bibfield{author}{\bibinfo{person}{Tao Zhou}, \bibinfo{person}{Gang Yan}, {and}
  \bibinfo{person}{Bing-Hong Wang}.} \bibinfo{year}{2005}\natexlab{}.
\newblock \showarticletitle{Maximal planar networks with large clustering
  coefficient and power-law degree distribution}.
\newblock \bibinfo{journal}{\emph{Physical Review E}} \bibinfo{volume}{71},
  \bibinfo{number}{4} (\bibinfo{year}{2005}), \bibinfo{pages}{046141}.
\newblock


\bibitem[Zhou and Nakhleh(2011)]%
        {zhou2011properties}
\bibfield{author}{\bibinfo{person}{Wanding Zhou} {and} \bibinfo{person}{Luay
  Nakhleh}.} \bibinfo{year}{2011}\natexlab{}.
\newblock \showarticletitle{Properties of metabolic graphs: biological
  organization or representation artifacts?}
\newblock \bibinfo{journal}{\emph{BMC Bioinformatics}} \bibinfo{volume}{12},
  \bibinfo{number}{1} (\bibinfo{year}{2011}), \bibinfo{pages}{1--12}.
\newblock


\end{thebibliography}
